\documentclass[12pt]{article}
\usepackage{amsmath}
\usepackage{mathrsfs}
\usepackage{amssymb}
\usepackage{color}
\usepackage{psfrag}
\usepackage{graphicx}
\usepackage{subfigure}
\usepackage{rotating}
\usepackage{cite}
\usepackage[colorlinks=true,linktocpage=true,linkcolor=blue,citecolor=blue]{hyperref}
\usepackage[section]{placeins}
\usepackage{amsfonts}
\usepackage{ifpdf}
\usepackage[toc,page]{appendix}
\usepackage{pifont}\usepackage{amsmath}
\usepackage[mathscr]{eucal}
\usepackage[all]{xy}
\usepackage{listing}
\usepackage{pdfpages}

\newcommand{\bea}{\begin{eqnarray}}
\newcommand{\eea}{\end{eqnarray}}
\newcommand{\be}{\begin{equation}}
\newcommand{\ee}{\end{equation}}

\newcommand{\vs}[1]{\vspace{#1 mm}}

\begin{document}
\topmargin 0pt
\oddsidemargin 0mm

\vspace{2mm}

\begin{center}

{\Large \bf {AdS black holes with higher derivative corrections in presence of string cloud}}

\vs{10}

{Tanay K. Dey$^*$ \footnote{E-mail: tanay.dey@gmail.com and tanay.d@smit.smu.edu.in} and Subir Mukhopadhyay$^\dagger$ \footnote{E-mail:  subirkm@gmail.com and smukhopadhyay@cus.ac.in}}

 \vspace{4mm}

{\em
$^*$Department of Physics, Sikkim Manipal Institute of Technology, Sikkim Manipal University,
Majitar, Rongpo, East Sikkim, Sikkim-737136, India.}

\vspace{4mm}

{\em
$^\dagger$ Department of Physics, Sikkim University, 6th Mile, Gangtok 737102.}

\end{center}

\vs{10}

\begin{abstract}
\noindent We consider asymptotically AdS black hole solutions in presence of string clouds and higher derivative corrections. We find that it admits three black hole solutions within a region of the parameter space. In order to examine the stabilities of these solutions we consider a thermodynamic analysis. Using holography, we have studied the quark-antiquark distance and binding energy in the dual gauge theory.
\end{abstract}

\newpage

\section{Introduction}\label{intro}
Study of strongly coupled gauge theories remain a challenge due to lack of appropriate systematic formulations and machineries. On that score, the AdS/CFT duality has been proved to be quite a useful route. This duality implies that $SU(N_c),\, {\cal N} = 4 $ SYM theory living on the boundary of the one higher dimensional space time in the limit of strong coupling is dual to a weakly coupled gravity theory in the higher dimensional space time and vice versa \cite{ Maldacena, Witten1, Witten2}. It has been found that there is a correspondence between the black hole configurations in the gravity theory and the dual boundary theory in its deconfined phase. On the other hand, the confined phase of the dual gauge theory is dual to pure AdS space time configuration. The  transition \cite{H-P} between the pure AdS configuration and the black hole configuration has been interpreted as equivalent to the  transition between the confined and deconfined phases of the dual gauge theory living on the boundary.

 A more realistic scenario is to consider a gauge theory with fundamental quarks and baryons, which are absent in $SU(N_c),\, {\cal N} = 4 $ Super Yang-Mills (SYM) theory. In order to incorporate fundamental quarks, \cite{Guen, Karch, Head, Big, Kumar} consider cloud of hanging strings whose end points are tracing out the contour of the loop at the asymptotic boundary. In the dual gauge theory living at the boundary, the end points of the string correspond to the quark and anti-quark, whereas the string itself corresponds to the gluons.

In view of this correspondence, clouds of strings are considered in \cite{shankha} in AdS black hole space time and it was found that in presence of this string clouds, the black hole configuration becomes a stable configuration with respect to its decay into the pure AdS. Furthermore in \cite{tanay1}, one of us observed that  this theory admits three different black hole solutions provided density of the string clouds lies below a certain value. The black hole solution having smallest horizon radius is called small black hole and the other two solutions are called medium and large depending on the sizes of their horizon radii. A thermodynamic analysis reveals that at lower (higher) values of  temperature, the stable configuration corresponds to the black hole solution, which has smallest (largest) radius. At a critical temperature, there occurs a Hawking-page like phase transition between these two black holes. In a similar model \cite{Yang, Yuan}, the dual phases of such black holes has been analysed in terms of the quark-antiquark($Q\bar Q$) separation and it is found that for large black hole there is an upper bound for the $Q\bar{Q}$ distance, while for small black hole it can increase indefinitely. Based on this observation, they concluded that the small black hole corresponds to the confined phase of the dual theory, while the deconfined phase corresponds to the large black hole. A similar study of the quark-antiquark distance has been carried out in \cite{tanay1} for the dual phases of small and large black hole solutions obtained with the string clouds.

In this work, we have examined how corrections in the above mentioned gravity theory\cite{shankha,tanay1} due to higher derivative terms, which amounts to considering subleading correction in `t Hooft coupling in the dual theory, modify different aspects of these solutions. In particular, we have explored the modifications of thermodynamics of black hole solutions and the quark-antiquark distance, that ensued due to higher derivative terms in the gravity theory. The particular higher derivative corrections, that we consider are the Gauss-Bonnet (GB) term and find that there is an upper bound of the coefficient of the GB term, beyond which the three different solutions cease to exist. We also find a Hawking-Page-like phase transition\cite{H-P}, which is similar to that obtained in absence of higher derivative corrections, as mentioned above. We have studied the Q$\bar Q$ distance, as well as the binding energy. We find the screening length of Q$\bar Q$ pair increases as the temperature decreases and though for the small black hole, it can go to a larger value compared to that of large black hole, there remains an upper bound. Therefore one can conclude that the present model does not admit a confined phase.  From the study of the binding energy, we find that it is smaller in small black hole background than that of the large back hole, showing  $Q\bar Q$ bound state is more stable in the former background. In particular, at high enough temperature, the Q$\bar Q$ distance vanishes showing that the Q$\bar Q$ bound state does not exist.

The paper is structured as follows. In section \ref{gravitydual}, a brief discussion of the black hole solution in presence of external string cloud is included. We then find the thermodynamical quantities and their behaviour in section \ref{thermo} and \ref{phase} respectively. Thereafter in section \ref{gaugetheory} we check the instability of $Q\bar Q$ bound state by studying the depth of the U-shape string hanging from boundary of the asymptotically AdS background. Finally we study  $Q\bar Q$ bound state and its binding energy at low temperature in section \ref{potential}. At the end we summarize our work in section \ref{conclude}.

\section{Black hole solution}\label{gravitydual}
We begin with a brief review of black hole solutions, which are asymptotically AdS in presence of string clouds. The string clouds corresponds to introducing strings in the background. They are extended from the boundary to  the horizon of the black hole or the center of the space time. According to the holography, this system represents certain states in the gauge theory in presence of a large number of quarks and has already been studied in \cite{tanay1}. In order to incorporate the effect of the subleading correction in the `t Hooft coupling we will add higher derivative terms in the Einstein gravity with string clouds. To be more specific, we will add the Gauss-Bonnet terms and will study their effects.

The $(4+1)$ dimensional gravitational action along with Gauss Bonnet terms is given by
\begin{equation}
\mathcal{S} = \frac{1}{16 \pi G_{5}} \int dx^{4+1} {\sqrt {-g}}( R - 2 \Lambda + \alpha L_{GB}) + S_m,
\label{totac}
\end{equation}
where $G_5$ represents the gravitational constant. $R$ is the Ricci scalar and $\Lambda$ is the cosmological constant. We write $g_{\mu\nu}$ as the space time metric tensor and $g$ represents its determinant. The term $$ L_{GB}=  R^2 + R_{\mu\nu\rho\sigma}R^{\mu\nu\rho\sigma} - 4 R_{\mu\nu}R^{\mu\nu}, $$ is the higher order correction term due to the quantum fields renormalization. This term is called Gauss-Bonnet correction in the Lagrangian. The $\alpha$ is the coefficient of the Gauss-Bonnet counterpart which play a crucial role in this theory.

In order to incorporate the string clouds we have added a further term, $S_{m}$. Since $S_m$ represents a large number of the strings its contribution can be written as,
\begin{equation}
S_m = -{\frac{1}{2}} \sum_i {\cal{T}}_i \int d^2\xi {\sqrt{-h}}
h^{\beta \gamma} \partial_\beta X^\mu \partial_\gamma X^\nu g_{\mu\nu}.
\label{matac}
\end{equation}
where we have integrated over world-sheet,
$h^{\beta \gamma}$ is world-sheet metric and $\beta, \gamma$ correspond to the world sheet coordinates. $S_m$ is the contribution of a large number of strings. We denote the tension of $i$'th string by ${\cal{T}}_i$.

We will assume that the strings are uniformly distributed over the three spatial directions and write the density as
\begin{equation}
a(x) = T \sum_i  \delta_i^3(x - X_i),~~~{\rm with} ~a > 0.
\label{denfun}
\end{equation}

With this assumption, one can show that the action (\ref{totac}) admits a black hole solution.  The metric tensor of this solution is \cite{Hers,Ghaff,Ghan},
\begin{equation}
ds^2  = -g_{tt}(r) dt^2 + g_{rr}(r) dr^2 + r^2 g_{ij} dx^i dx^j.
\label{genmet}
\end{equation}
Here $g_{ij}$ is the metric on the $(4-1)$ dimensional boundary and
\begin{equation}
g_{tt}(r) = 1 +\frac{r^2}{4 \alpha}\Big(1 - \sqrt{1+\frac{32 \alpha M}{r^4} - \frac{8 \alpha}{l^2} + \frac{16 a \alpha}{3 r^3}}\Big) = \frac{1}{g_{rr}},
\label{comp}
\end{equation}
where $l$ corresponds to the AdS radius and related to cosmological constant via the relation, $\Lambda = - \frac{6}{l^2}$.  $M$ is the constant of integration, which represents the mass of the solution.

The solution is singular at $r = 0$, which is cloaked by a horizon located at $r_+$, which can be obtained by setting $g_{tt}(r_+)=0$,
\begin{equation}
1 +\frac{r_+^2}{4 \alpha}\Big(1 - \sqrt{1+\frac{32 \alpha M}{r_+^4} - \frac{8 \alpha}{l^2} + \frac{16 a \alpha}{3 r_+^3}}\Big) = 0.
\label{hor}
\end{equation}
 To get the real solution  of the above equation, the discriminant should remain positive which implies following condition,
\begin{equation}
r^4_{bh}\Big(1-\frac{8\alpha}{l^2}\Big) + 16 \alpha\Big(2 M + \frac{a r_{bh}}{3}\Big) = 0,
\label{mas}
\end{equation}
to be imposed on the maximum value of the horizon radius.
We can trade the ADM mass of the black hole solution for the radius of the horizon using the following relation,
\begin{equation}
M = \frac{3r_+^4 + 3 l^2 r_+^2 -2 a l^2 r_+ + 6 l^2 \alpha}{12 l^2}.
\label{mas}
\end{equation}

This solution is asymptotically AdS and it is necessary to analyse the stability of this solution towards a decay into the pure AdS. We will consider a thermodynamic analysis of it in the next section. As we will see, for a given temperature it admits three different kinds of black holes and we will study the relative stability of these solutions.

\section{Thermodynamics}\label{thermo}

Black holes can be considered
as thermodynamic systems and various thermodynamical quantities associated to black hole maintain a similar type of thermodynamical law as thermodynamic system followed. In the present section we will consider the solutions and their different thermodynamic aspects. In what follows, we set the term related to gravitational constant, $16 \pi G_5 =1$ and also consider unit volume of the 3-sphere.

\begin{itemize}

\item {\bf Temperature:} To begin with the black hole temperature is given by,
\begin{equation}
T = \frac{1}{4 \pi}\frac{d g_{tt}}{dr}|_{r = r_+}
= \frac{ 6 r_+^3 + 3l^2r_+ - a l^2}{6 \pi l^2(r_+^2 + 4 \alpha)},
\label{flattemp}
\end{equation}
This gives rise to a cubic equation in $r_+$ and for a given temperature, generically, we can expect three solutions for $r_+$.

\item {\bf Entropy:} Assuming that the first law of thermodynamics is satisfied by the black hole, we obtain the following expression for the entropy,
\begin{equation}
S = \int T^{-1} dM = \pi\Big(\frac{r_+^3}{3}+ 4 \alpha r_+\Big).
\end{equation}

\item {\bf Specific heat:} Specific heat can easily be obtained by using the formula $C= T \frac{dS}{dT}$ and is given by,
\begin{equation}
C = \frac{\partial M}{\partial T} =
\frac{\pi\big(r_+^2 + 4 \alpha\big)^2 \Big[6 r_+^3 + 3 l^2 r_+ - al^2\Big]}{6r_+^2\big(r_+^2 + 12 \alpha\big) + 2 a l^2 r_+ -3l^2\big(r_+^2 -4 \alpha\big)}.
\label{sh}
\end{equation}

\item{\bf Helmholtz free energy:} Considering the energy of the black hole $E$ to be equal to its mass we get the following form of the Helmholtz free energy, $F=E-TS$ for these black holes.
\begin{equation}\label{fenergy}
F=E-TS= -\frac{3 r_+^6 - 3 l^2 r_+^4 + 108 \alpha r_+^4 + 4 a l^2 r_+^3 + 18\alpha l^2 r_+^2 -72 \alpha^2 l^2}{36 l^2 (r_+^2 + 4 \alpha)}.
\end{equation}

\item{\bf Landau function:} In order to study the transitions between different phases, we have constructed the Landau function computed around the critical point, where radius of the horizon plays the role of the order parameter. The construction goes as follows; First we consider a function $G$ as a power series in the horizon radius $r_+$,
\begin{equation}
G(T, r_+) = \alpha_0  r_+^0 + \alpha_1  r_+^1 + \alpha_2  r_+^2 + \alpha_3  r_+^3 +\alpha_4  r_+^4 + \cdot \cdot \cdot,
\end{equation}
where $\alpha_i$ are functions of temperature $T$ and are chosen so as to satisfy the following conditions:

\item Up to a certain temperature $T_{min}$, $G$ will have only one minimum at $ r_{+} = r_{+1}$ with $r_{+1} \ge 0$.

\item Once the temperature is above $T_{min}$ another minimum appears at $ r_{+3}$. Minima at $r_+ = r_{+1}$ and $r_+ = r_{+3}$ must be separated by a maximum at $r_+ = r_{+2}$. We demand that at high temperature, the second minimum is globally stable. In order to achieve this, we can consider only up to the quartic power of the order parameter and neglect all the terms, which are higher order in the order parameter. Therefore, the Landau function can be written as;
\begin{equation}
G(T, r_+) =\alpha_0 + \alpha_1  r_+^1 + \alpha_2  r_+^2 + \alpha_3  r_+^3 +\alpha_4  r_+^4.
\end{equation}

\item In order to determine these five $\alpha_i$ we impose the following five conditions.
    \begin{itemize}
    \item At the three extreme points the function should satisfy $ \frac{\partial G}{\partial r_+}|_{r_1, r_2, r_3}= 0$,
    \item Expression of the temperature can be obtained from the condition of extrema of this function.
    \item Finally we get back the expression of free energy of the system by substituting temperature in the Landau function.
\end{itemize}
\end{itemize}
    From the above five conditions we can compute the constants $(\alpha_0, \alpha_1, \alpha_2, \alpha_3,\alpha_4)$.  The Landau function, when written in terms of temperature and the order parameter, reduces to the following expression,
\begin{equation}\label{landau}
G= \frac{3 r_+^4 - 4\pi l^2 r_+^3 T + 3 l^2 r_+^2 - 48 \pi \alpha l^2 r_+ T - 2 a l^2 r_+ + 6 \alpha \l^2}{12 l^2}.
\end{equation}

An analysis of these thermodynamic quantities of the black hole and their dependence on various parameters of the model reveals a complex thermodynamic phase structure. This will enable us to study the issue of thermodynamic stability of the various solution admitted by the present model.

\section{Phases of black hole}\label{phase}

In order to explore the phase structures of the solutions, we will examine the behaviour of the various thermodynamic quantities with respect to variation of different parameters. For the sake of simplicity of notation, we scale a,  $\alpha$ and the horizon radius with appropriate powers of l to render them dimensionless as follows:
\be \bar a = \frac{a}{l}, \quad \bar \alpha = \frac{\alpha}{l^2}\quad\text{and}\quad  \bar r = \frac{r_+}{l}.\ee

We have introduced the temperature for the black hole solutions in (\ref{flattemp}), which can be expressed as
\begin{equation}
\bar T=\frac{6\bar r^3 + 3 \bar r - \bar a}{6 \pi l (\bar r^2 + 4 \bar \alpha)}.
\end{equation}
It shows that for a fixed temperature, the scaled horizon radius ${\bar r}$ satisfies a cubic equation. If the cubic equation admits three real and positive solutions there are three black holes or it will be only one black hole solution.

\begin{figure}[h]
\begin{center}
\begin{psfrags}
\psfrag{T}[][]{$\bar T$}
\psfrag{r}[][]{$\bar r$}
\psfrag{a}[][]{$\bar a$}
	\mbox{
	\subfigure[]{\includegraphics[width=7.5cm]{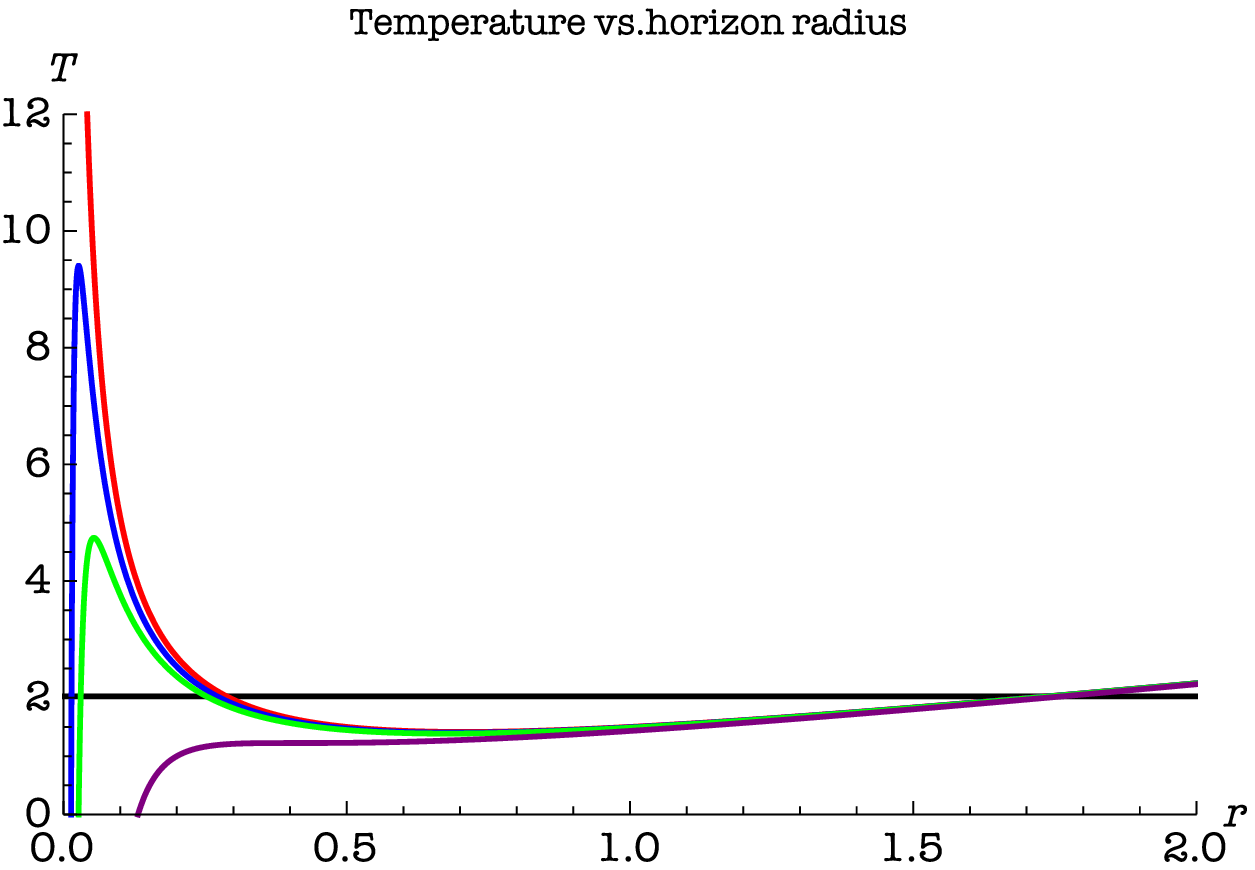}}
\quad
	\subfigure[]{\includegraphics[width=7.5cm]{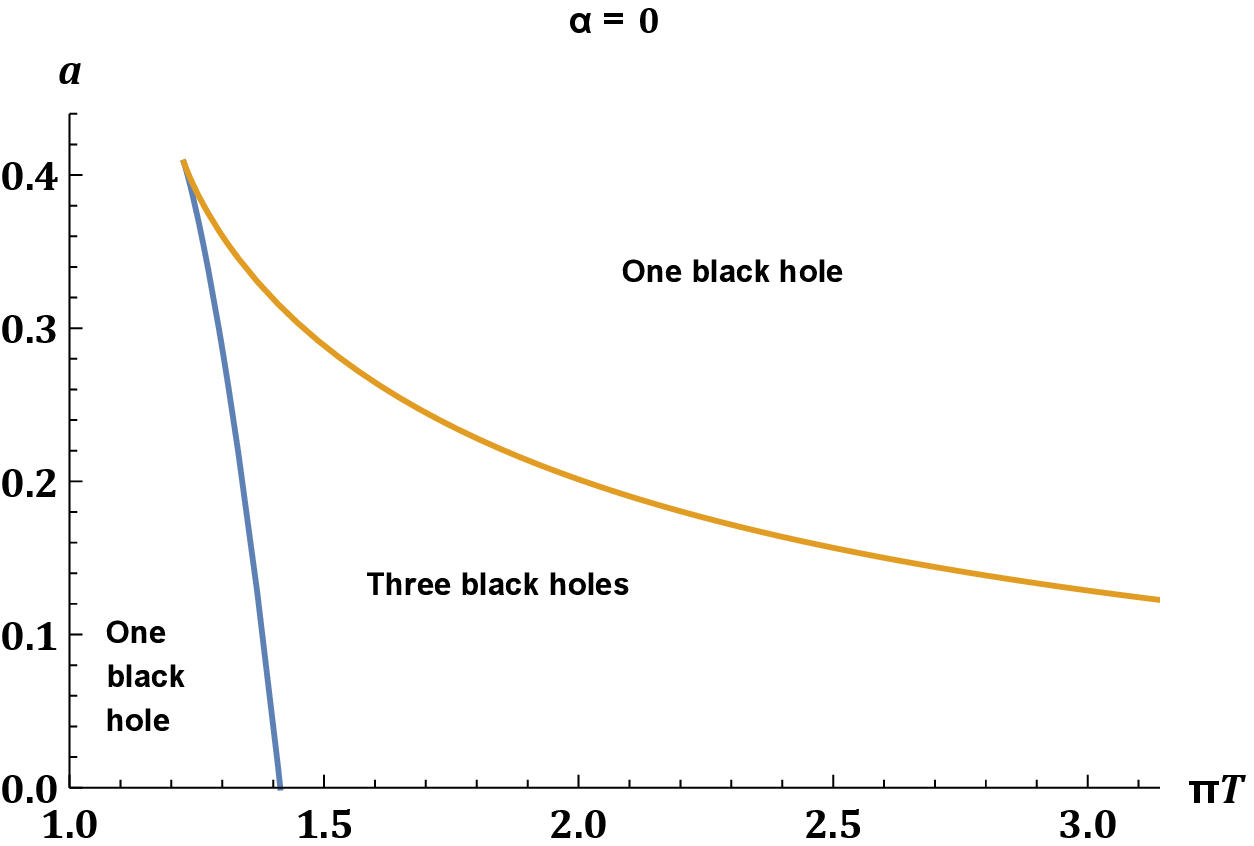}}}
\end{psfrags}
\end{center}
\caption{Both the graphs are for $\bar\alpha=0$.(a) Temperature $\bar T $ vs. scaled horizon radius $\bar r$ and (b) string cloud density $\bar a$ vs. temperature $\pi \bar T$.}
\label{tempar}
\end{figure}

The discriminant of the cubic equation is given by,
\be
\triangle =-\Big [ 16 \bar\alpha (\pi \bar T)^4 + \frac{2 {\bar a}}{3} (\pi \bar T)^3 + \big(432 \bar\alpha^2 - 36 \bar\alpha -\frac{1}
{4}\big) (\pi \bar T)^2 - 9 \bar a \big(\frac{1}{6} - 4 \bar\alpha \big) (\pi \bar T) + \frac{3}{4} {\bar a}^2 + \frac{1}{2}\Big ].
      \ee
In the above equation and in the following plots we have set $l=1$. In the region of the parameter space where $\triangle >0$, it admits three real solutions and otherwise there is only one real solution. First we have plotted the region for $\triangle > 0$ for $\bar\alpha = 0$ in figure \ref{tempar}.

In the left subfigure of figure \ref{tempar}, we consider a set of values of $\bar a$, the string cloud density and plotted temperature against $\bar r$, the radius of the horizon scaled appropriately. We observe that for $\bar a =0$(red curve), there are two black hole solution for any temperature greater than a minimum temperature. However once $\bar a$ gets some positive value but below a critical value $\bar a = \bar a_c = \frac{1}{\sqrt {6}}=.408$(for details see \cite{tanay1})there are certain range of temperature where three black hole solutions exist. Outside this range only one black hole solution exist. Once $\bar a$ is above its critical value (purple curve), for all values of temperature we get one black hole solution.

 In right subfigure of figure \ref{tempar} we plot string cloud density $\bar a$ against $\pi \bar T$.  For a given value of string cloud density $0<\bar a < \bar a_c$ there are two values of temperatures $\bar T_1$ and $\bar T_2$, such that the three black hole solutions exist in the region between $\bar T_1$ and $\bar T_2$. For $\bar a=0$, $\pi \bar T_1= \sqrt{2}$ and $\bar T_2$ is extended up to infinity and for $\bar T> \sqrt{2}/\pi$, we get only two black hole solutions, which cease to exist as we decrease the temperature below $\sqrt{2}/\pi$.

As we turn on the string clouds, its density $\bar a$ becomes non-zero and $\bar T_2$ becomes finite. As the density of the string clouds $\bar a$ increases more and more, the upper bound for three solutions $\bar T_2$ keeps on decreasing. On the other hand, the lower bound $\bar T_1$ also decreases but at a much slower rate. Finally, when $\bar a$ reaches a critical value, $\bar a = \bar a_c$, $\bar T_1$ and $\bar T_2$ meet, indicating the merger of all the three black hole solutions. For $\bar a > \bar a_c$, there exists only a single black hole solution.

\begin{figure}[h]
 \begin{center}
\begin{psfrags}
\psfrag{T}[][]{$\bar T$}
\psfrag{\alpha}[][]{$\bar \alpha$}
\psfrag{a}[][]{$\bar a$}
\includegraphics[width=8.5cm]{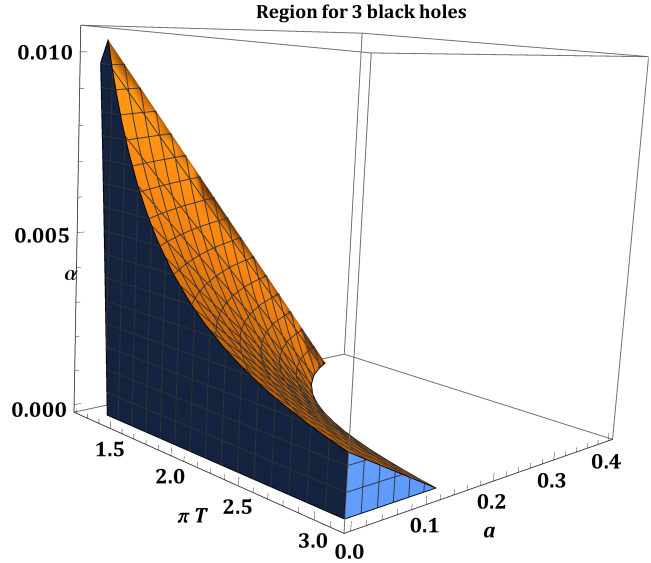}
\end{psfrags}
\end{center}
\caption{Temperature vs string cloud density with higher derivative correction.}
\label{discr}
\end{figure}

 Once we turn on the higher derivative correction $\bar \alpha$, the region of the three dimensional parameter space corresponding to $\bar T$, $\bar a$ and $\bar \alpha$, which admits three black hole solutions are shown in the figure \ref{discr}. As $\bar \alpha$ increases, the upper bound of the temperature for the existence of three black hole solution $\bar T_2$ becomes finite. For example, for $\bar \alpha = . 001$ and $\bar a=0$, $\bar T_2$ becomes approximately $3/\pi$. $\bar T_2$ keeps on decreasing with increase of $\bar \alpha$. The critical value of $\bar a$ denoted by $\bar a_c$, where the two bounds $\bar T_1$ and $\bar T_2$ meet, keeps on decreasing as $\bar \alpha$ increases and vanishes at around $\bar \alpha_c = .01386$. For $\bar \alpha > \bar\alpha_c$ there will be only a single black hole solution.

\begin{figure}[h]
\begin{center}
\begin{psfrags}
\psfrag{T}[][]{$\bar T$}
\psfrag{r}[][]{$\bar r$}
\mbox{\subfigure[]{\includegraphics[width=7.5 cm]{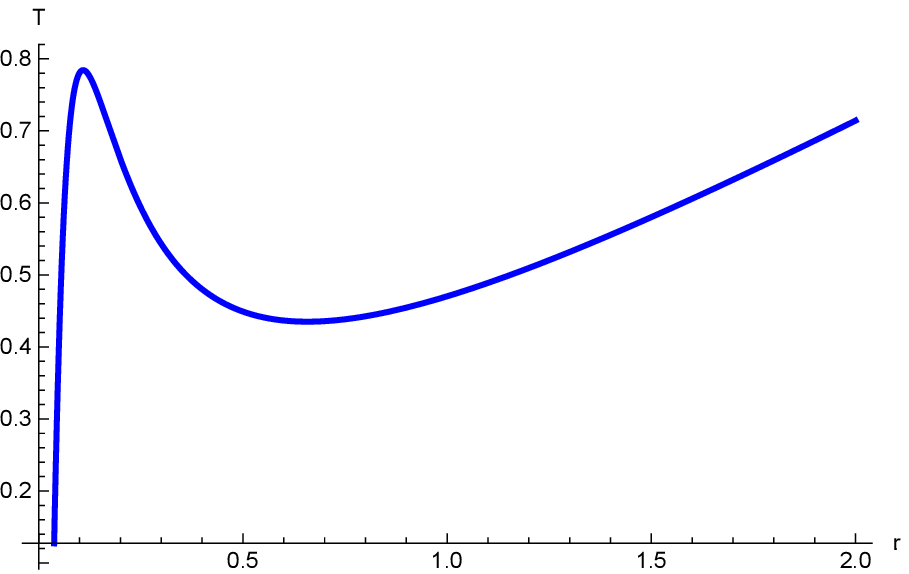}}
\quad
\subfigure[]{\includegraphics[width=7.5 cm]{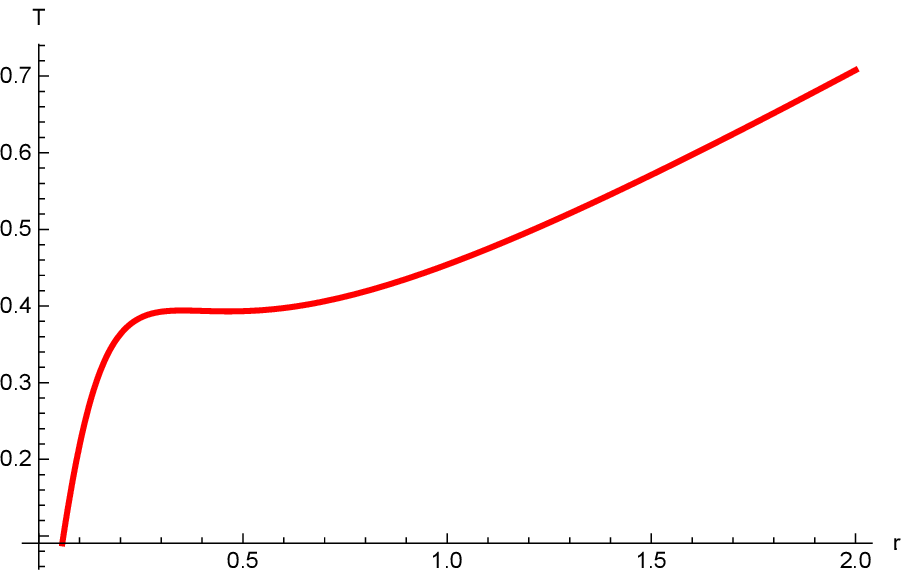}}}
\caption{$\bar T$ vs. ${\bar r}$ plots. (a) $\bar a=0.1,  \bar \alpha = 0.001$ and (b) $\bar a=0.1, \bar \alpha = 0.01$.
}
\label{temp}
\end{psfrags}
\end{center}
\end{figure}

In the figure \ref{temp} we have plotted temperature with respect to $\bar r $ for two specific values of string cloud density $\bar a$  and Gauss-Bonnet coupling $\bar \alpha $  . It shows that at a particular temperature, the horizon radii of these three black hole solutions are of different sizes and we call them small, medium and large depending on their horizon radii.
For $\bar T < \bar T_1$ only a single black hole exists, whose radius decreases as the temperature decreases. This single black hole persists even at zero temperature, where the radius reaches its minimum value, which can be expanded in a power series of $\bar a$,
\begin{equation}
\bar r_{0} = \frac{\bar a}{3}  + \mathcal{O}(\bar a^2).
\label{rmin}
\end{equation}

Having established the different black hole solutions from the study of their temperatures, we will enquire about their thermodynamic stability in the respective domains of the parameter space. For this purpose, we will examine the three different thermodynamic quantities: the specific heat, the Helmholtz free energy and the Landau function, that we have introduced in the earlier section.

We begin with the specific heat. The expression for the specific heat of the black hole solutions in terms of the three dimensionless parameters turns out to be;
\begin{equation}
\bar C=\frac{\pi l^3 (4 \bar\alpha + \bar r^2)^2 [6\bar r^3 + 3  \bar r - \bar a]}{6\bar r^2(\bar r^2 + 12 \bar \alpha) - 3 (\bar r^2 - 4 \bar \alpha) + 2\bar a \bar r}.
\end{equation}

\begin{figure}[h]
\begin{center}
\begin{psfrags}
\psfrag{r}[][]{$\bar r$}
\psfrag{T}[][]{$\bar T$}
\psfrag{C}[][]{$\bar C$}
	\mbox{
	\subfigure[]{\includegraphics[width=7.5cm]{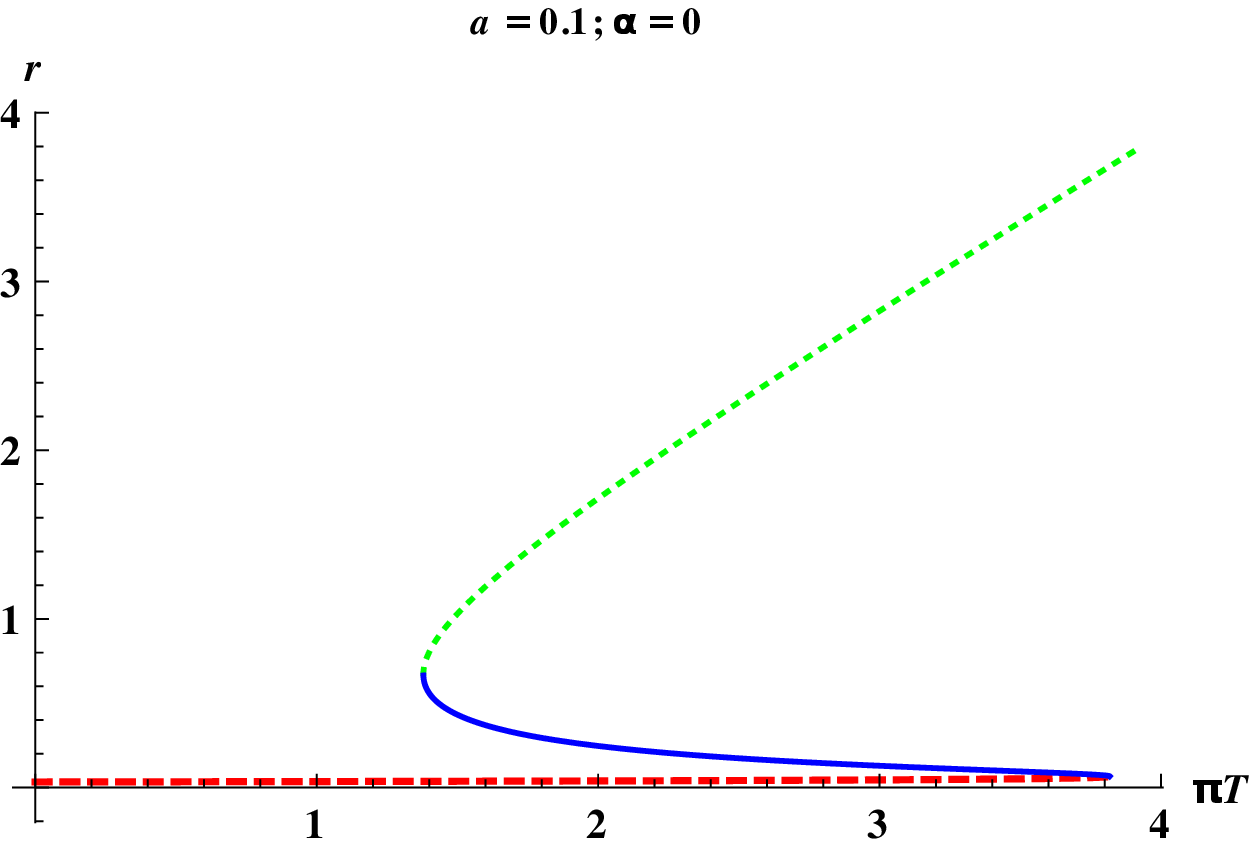}}
\quad
	\subfigure[]{\includegraphics[width=7.5cm]{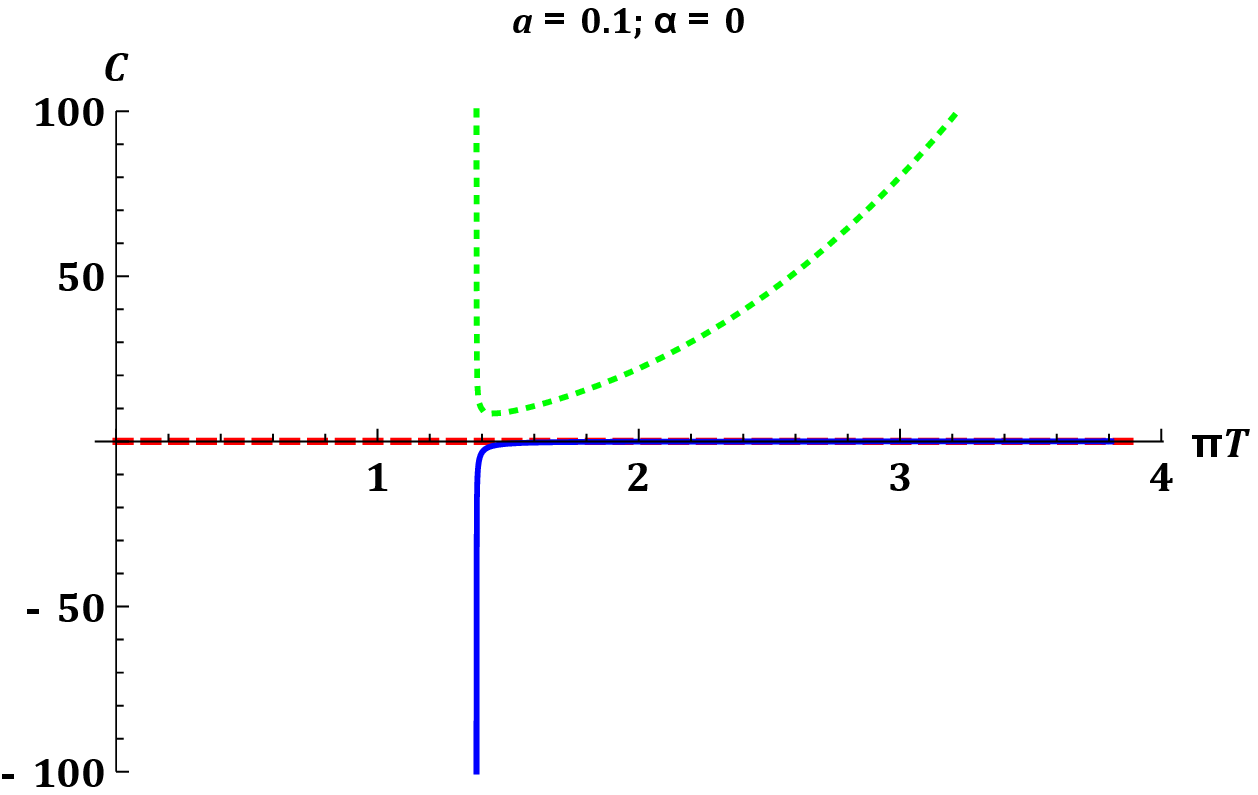}}}
\end{psfrags}
\end{center}
\caption{(a) $\bar r$ vs $\pi \bar T$ Plot (b) $\bar C$ vs. $\pi \bar T$  with $\bar \alpha=0$, $\bar a=0.1$ for black holes with horizon radii small (red), medium (blue) and large(green) . }
\label{spht}
\end{figure}

We have plotted the specific heat vs. temperature in figure \ref{spht}(b) along with a plot of the horizon radii vs. temperature in figure \ref{spht}(a) and we have chosen  $\bar\alpha=0$, $\bar a=0.1$. In these figures, the red dashed line represents the small black hole which exists up to $\bar T < \bar T_2$ and as one can observe from figure \ref{spht}(b), its specific heat remains positive (though small) throughout the range where it exists. The blue line represents the medium black hole, which exists between $\bar T_1$ and $\bar T_2$ and the figure \ref{spht}(b) shows that  the specific heat is negative leading to thermodynamic instability of the solution. The green, dotted line represents the large black hole. It starts off its existence from $\bar T = \bar T_1$ and persists for the entire range of temperature greater than $\bar T_1$. Once temperature is above $\bar T_2$ it represents the single black hole. As one can observe in \ref{spht}(b), it has a positive specific heat, showing it is thermodynamically stable. The behaviour of specific heat at the temperature, where unstable medium black hole meets the black holes with large horizon radius, indicates that the medium black hole decays at $\bar T_1$ through a first order phase transition.

\begin{figure}[h]
\begin{center}
\begin{psfrags}
\psfrag{T}[][]{$\bar T$}
\psfrag{r}[][]{$\bar r$}
	\mbox{
	\subfigure[]{\includegraphics[width=7.5cm]{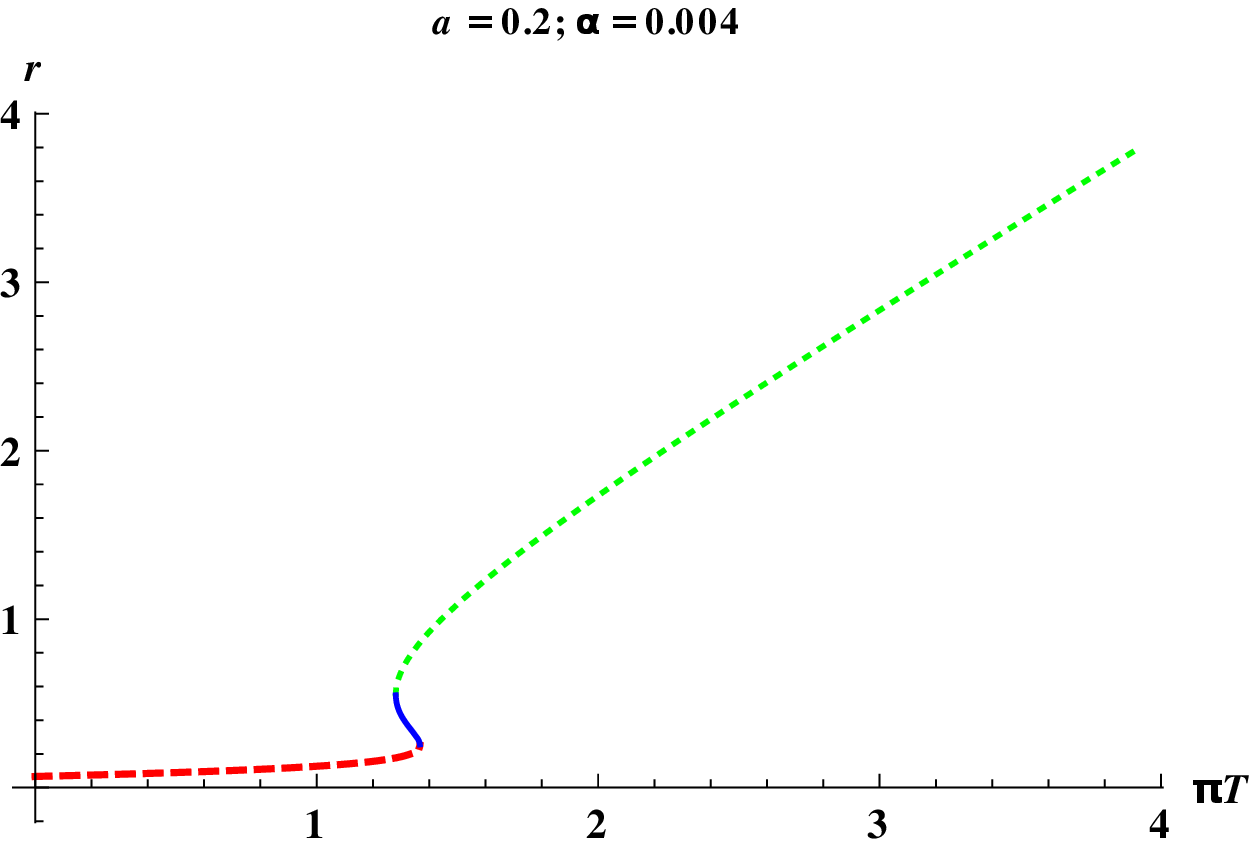}}
\quad
	\subfigure[]{\includegraphics[width=7.5cm]{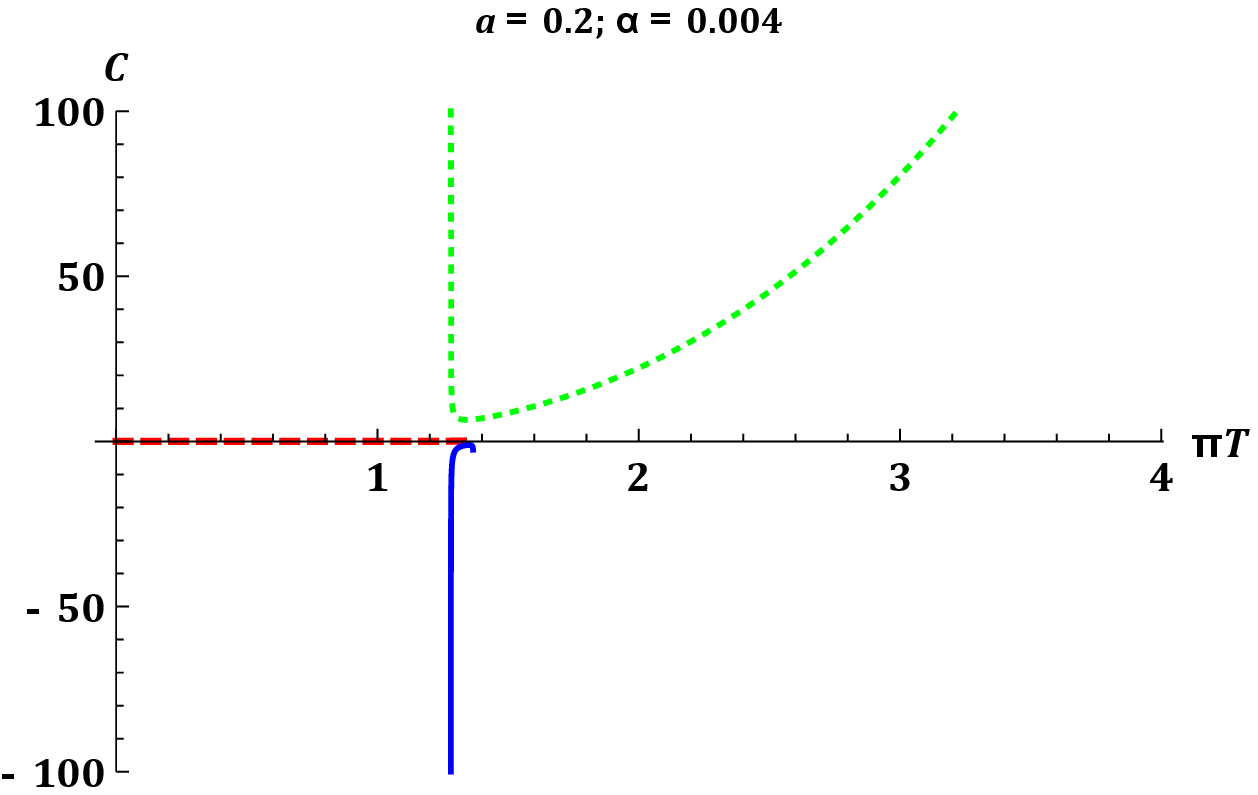}}}
\end{psfrags}
\end{center}
\caption{(a) $\bar r$ vs $\pi \bar T$ Plot (b) $\bar C$ vs. $\pi \bar T$  with $\bar \alpha=0.004$, $\bar a=0.2$ for black holes with horizon radii small (red), medium (blue) and large(green) . }
\label{sphtalpha}
\end{figure}

Similar diagrams has been plotted in figure \ref{sphtalpha}(a) and figure \ref{sphtalpha}(b) for $\bar \alpha=.004$, $\bar a=0.2$. As discussed earlier, due to turning on of the higher derivative correction, $\bar T_2$ has been reduced and approached $\bar T_1$ resulting the range of the temperature for the existence of the medium black hole smaller. The specific heat shows qualitatively similar features.

Next, we analyse the Helmholtz free energy, which  can be written as;
\begin{equation}
\bar F=-\frac{l^2[3 \bar r^6 - 3\bar r^4 +108 \bar \alpha \bar r^4 + 4 \bar a \bar r^3 + 18 \bar \alpha \bar r^2 -72 \bar \alpha^2]}{36 (\bar r^2 + 4 \bar\alpha)}.
\end{equation}

\begin{figure}[h]
\begin{center}
\begin{psfrags}
\psfrag{F}[][]{$\bar F$}
\psfrag{T}[][]{$\bar T$}
	\mbox{
	\subfigure[]{\includegraphics[width=7.5cm]{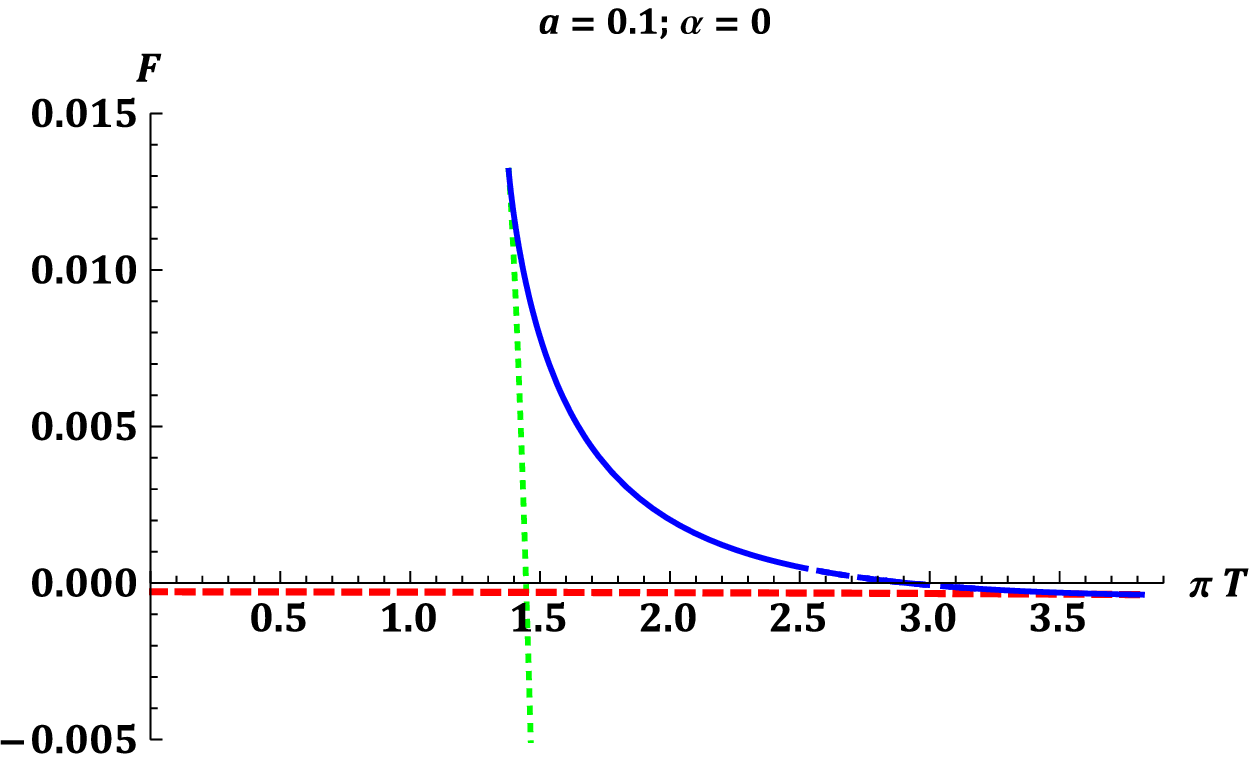}}
\quad
	\subfigure[]{\includegraphics[width=7.5cm]{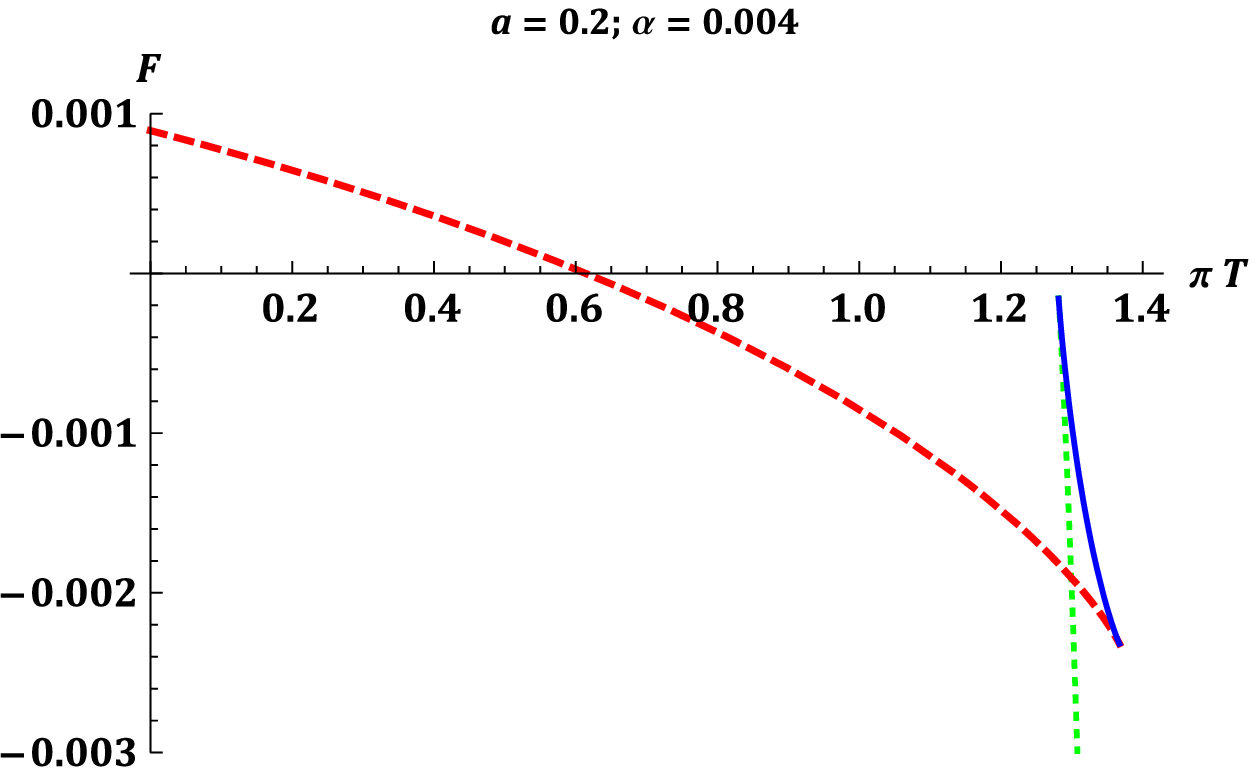}}}
\end{psfrags}
\end{center}
\caption{Free energy is plotted against temperature: black holes with horizon radii small, medium and large are plotted in red, blue and green. (a) $\bar \alpha=0$, $\bar a=0.1$(b)  $\bar\alpha=0.004$, $\bar a=0.2$.}
\label{FEbh}
\end{figure}

In the figure \ref{FEbh}(a), we have plotted the free energy vs. the temperature for $\bar \alpha=0$, $\bar a=0.1$. Since for our choice of parameters, the free energy of the small black hole is always less than that of AdS configuration,  for small temperature the small black hole remains thermodynamically stable till $\bar T < \bar T_1$, which is consistent with the analysis of the specific heat. As $\bar T$ reaches $\bar T_1$, the other two black holes appear. At any temperature, a comparison of free energies shows the black hole with intermediate horizon radius is not thermodynamically favoured. For $\bar T > \bar T_c$, a critical temperature which is less than $\bar T_2$, the small black hole has a free energy greater than that of the large black hole and therefore, the large black hole continues to be the thermodynamically favoured configuration for the range of temperature $\bar T > \bar T_c$.

In the figure \ref{FEbh}(b), we have plotted the free energy vs. the temperature for $\bar\alpha=.001$, $\bar a=0.2$. One noticeable difference that happens here is that the free energy for the small black hole is positive up to a certain temperature and then it becomes negative. On the other hand the free energy for the medium and large black holes is negative for any temperature. Except these differences other behaviour are same as previous case.

\begin{figure}[h]
\begin{center}
\begin{psfrags}
\psfrag{F}[][]{$\bar F$}
\psfrag{r}[][]{$\bar r$}
\mbox{\subfigure[]{\includegraphics[width=7.5 cm]{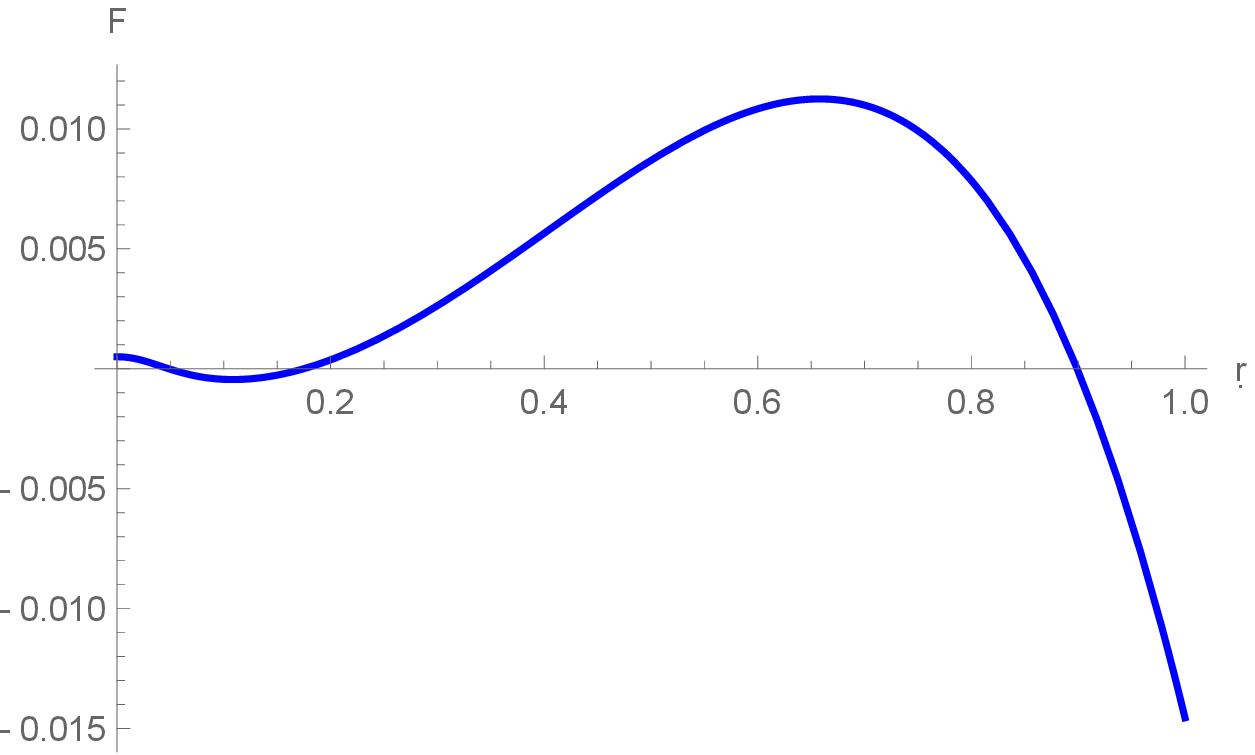}}
\quad
\subfigure[]{\includegraphics[width=7.5 cm]{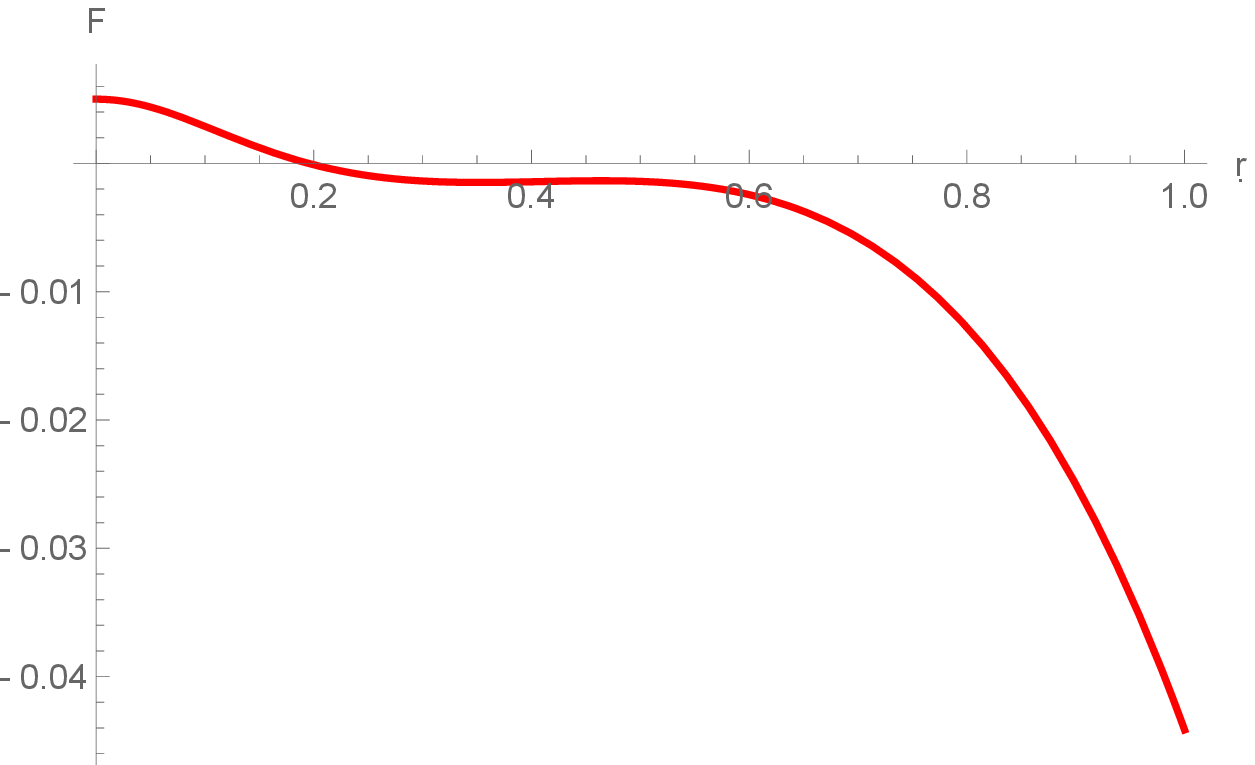}}}
\end{psfrags}
\caption{Free energy is plotted against horizon radius: (a) $\bar a=0.1$ and  $\bar \alpha = 0.001 $ and plot (b) is for $\bar a=0.1$, $\bar \alpha = 0.01 \,{\rm and}\, l=1$.}
\label{freeenergy}
\end{center}
\end{figure}

In the figure \ref{freeenergy}, we have plotted the free energy vs. the scaled horizon radius for two different sets: (a) $\bar a=0.1$, $ \bar \alpha = 0.001$ and (b) $\bar a=0.1$, $ \bar \alpha = 0.01$.  As one can observe from these plots at $\bar r = 0$ the free energy is positive. As the radius of the black hole increases, the free energy decreases and becomes minimum. As we increase ${\bar r}$ further, the free energy starts increasing, becomes positive and reaches a maximum value at a certain radius. For further increase of the horizon radius, the free energy becomes negative and keeps on decreasing. We can associate again free energy of $\bar r = 0$ as free energy of AdS configuration, the local minima of the free energy as the small black hole free energy and the negative free energy for the large $\bar r$ is related to the large black hole. The maxima in the free energy corresponds to the unstable black hole. From the consideration of free energy, AdS configuration is not a stable configuration and small and large black hole can be stable configuration only.

We consider the Landau function which is constructed around the critical point to investigate about stability of different solutions and is given by:
\begin{equation}
\bar G=\frac{3 l^2 \bar r^4 - 4\pi l^3 \bar r^3 T + 3 l^2 \bar r^2 - 2 l^2 \bar a \bar r - 48 \pi l^3 \bar \alpha \bar r T + 6 \bar\alpha l^2}{12}.
\end{equation}
\begin{figure}[h]
\begin{center}
\begin{psfrags}
\psfrag{G}[][]{$\bar G$}
\psfrag{r}[][]{$\bar r$}
\mbox{\subfigure[]{\includegraphics[width=7.5 cm]{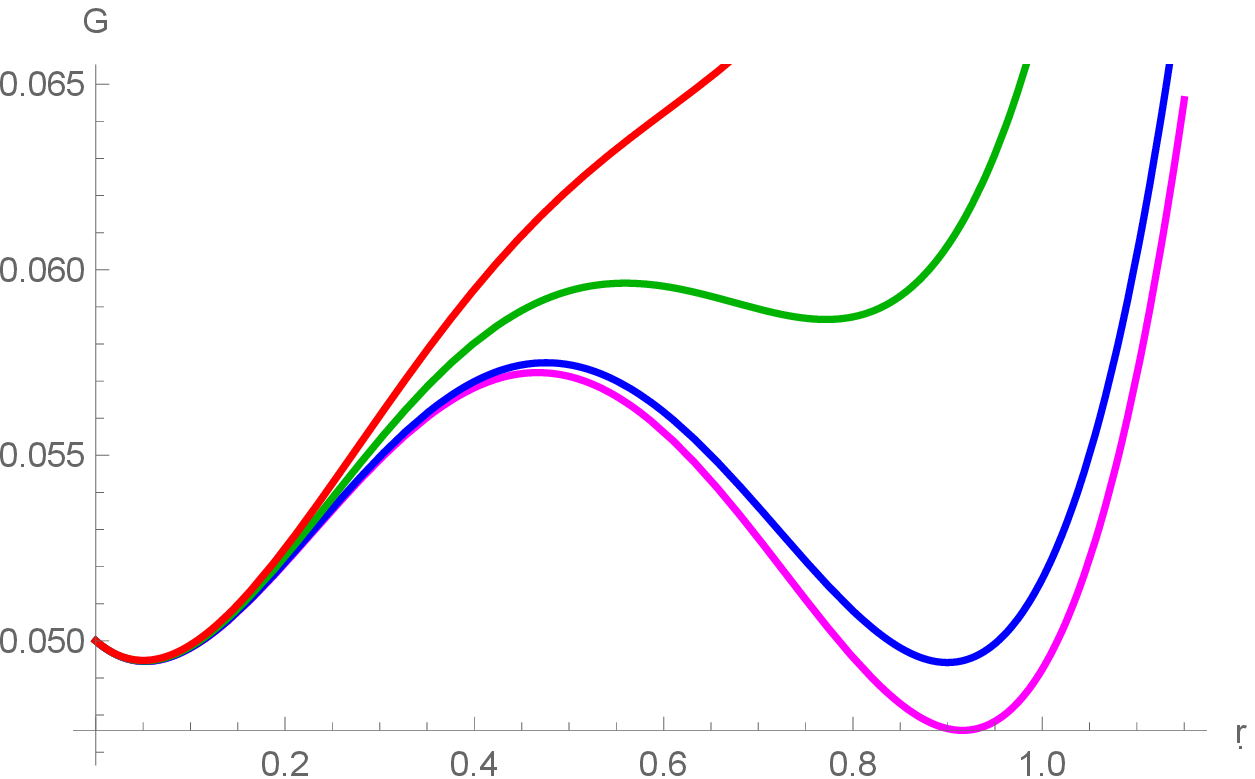}}
\quad
\subfigure[]{\includegraphics[width=7.5 cm]{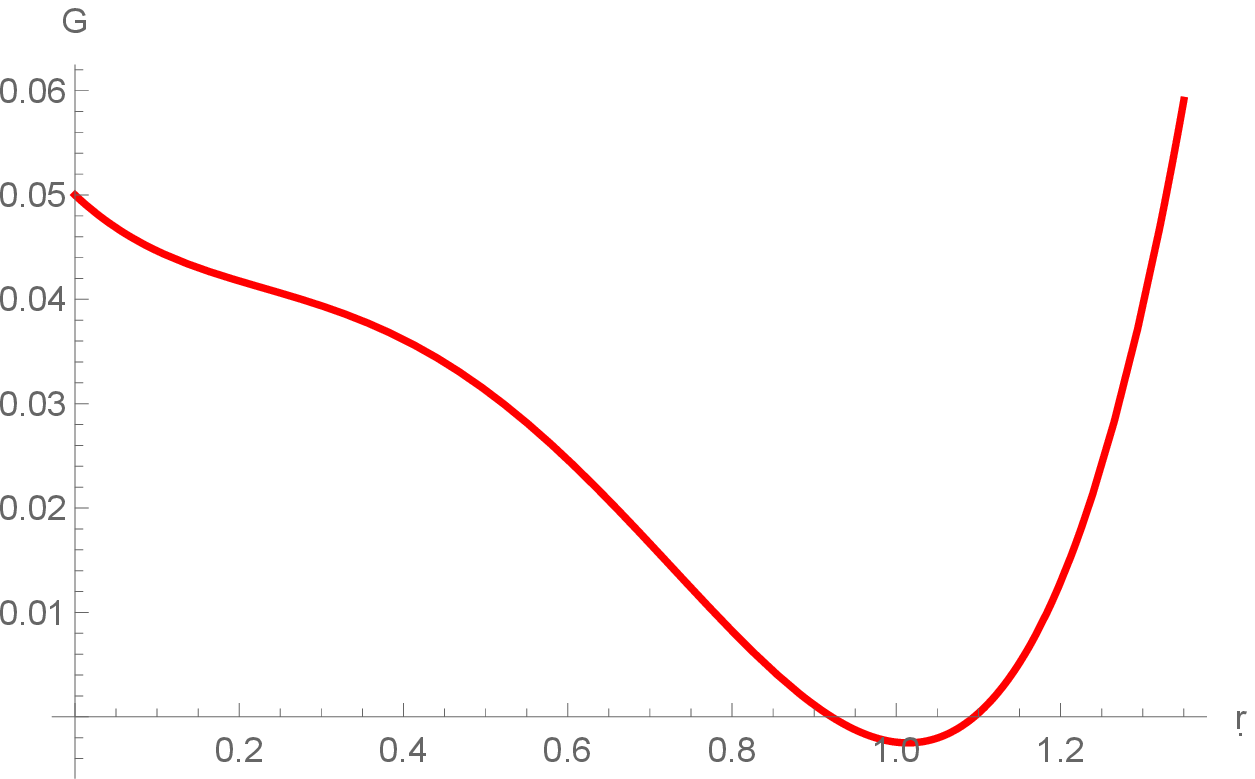}}}
\end{psfrags}
\caption{Landau function vs. $\bar{r}$ for four different temperatures: $\bar T_0 =0.42$ (red) , $\bar T_1 = 0.44$ (green) , $\bar T_c = 0.4545$ (blue) $\bar T_{c1} = 0.4568$ (pink) ;  $\bar T_0 < \bar T_1 < \bar T_c < \bar T_{c1}$ (a)$\bar a=0.1$ and  $ \bar \alpha = 0.001$ (b) is for $\bar a=0.1$, $\bar \alpha = 0.01 \,{\rm and}\, l=1$. .}
\label{gibbs}
\end{center}
\end{figure}
We consider four different temperatures and have plotted Landau function with respect to scaled horizon in subfigure \ref{gibbs}(a) with $\bar a=0.1$ and $ \bar \alpha = 0.001$. The temperatures we consider are $\bar T_0 =0.42$(red curve),$\bar T_1 = 0.44$(green curve), $\bar T_c = 0.4545$(blue curve) and $\bar T_{c1} = 0.4568$(pink curve) such that $\bar T_0 < \bar T_1 < \bar T_c < \bar T_{c1}$.  For temperature $\bar T_0$, only one minimum of Landau function at finite $\bar r$ exists, that corresponds to the black hole solution with small horizon radius. With rise of temperature two more extrema nucleates at temperature $T_1$.  The maxima of the Landau function corresponds to unstable black hole and minima corresponds to the large black hole. As we increase the temperature further at temperature $T_c$ the Landau function associated with the black holes having smallest and largest radii become equal. Above $T_c$, the Landau function for large black hole becomes lower than that of the small black hole indicating stability of the small black hole at low temperature and of the large black hole at high temperature. It also indicates that there will be a transition between the small and large black hole at $T=T_c$ similar to Hawking-Page phase transition. In subfigure \ref{gibbs}(b), we plot the Landau function against scaled horizon radius for $\bar a=0.1$ and $ \bar \alpha = 0.01$. We observed that there is only one minima at finite $\bar r$ corresponds to existence of  one stable black hole solution from which one may conclude that pure AdS is not a stable configuration at any temperature.

\section{Quark-antiquark distance}\label{gaugetheory}

As pointed out in the introduction, it is interesting to examine the phase of the dual model with reference to the confinement. For that purpose, we consider a string configuration having its endpoints on the boundary and elongated in the bulk. There may be two such configurations: A straight string from boundary to horizon or a U-shaped string where the string is hanging from the boundary, with its tip located at $u_0$, but not touching the horizon.

 We will denote the distance between the two endpoints by $L$, which is the quark-antiquark distance associated with the bound state in the boundary theory. There is an upper bound $L_s$ of the quark-antiquark separation $L$.  
 Every $L$, $0<L<L_s$ corresponds two U-shaped strings with tip $u_0$ being nearer to and farther from the horizon. As $L$ increases, these two configurations approach each other and they merge at $L=L_s$, when the quark-antiquark distance reaches the screening length.

Quark-antiquark distance for a bound state can be obtained by considering a probe string in the black hole spacetime. The Nambu-Goto world sheet action for such an open string is,
\begin{equation}{\label{ng}}
S_{NG} = - \frac{1}{2 \pi \alpha^\prime } \int d^2\sigma \sqrt{- det (h_{\beta \gamma})},
\end{equation}
where $\frac{1}{2 \pi \alpha^\prime}$ is the string tension. The string tension is related to $\lambda$, the `t Hooft coupling in the dual SYM gauge theory \cite{Ewerz:2016zsx},
\begin{equation}
\sqrt{\lambda} = \frac{l^2}{\alpha^\prime}.
\end{equation}
The present case being a bottom-up approach, we do not have a precise description of the dual boundary theory. Nevertheless we can expect that the $\lambda$ will play an analogous role. We will assume that the above Nambu-Goto action is a valid action for our problem and corrections ensuing from coupling to the further bulk fields will not qualitatively modify the essential result.

For convenience, we will use $u$ rather than $r$, where $r= \frac{l^2}{u}$ so that u = 0 is the boundary and the horizon occurs at $u_h$. The metric tensor of the black hole solution of the equation (\ref{genmet}) reduces to the form as,
 \begin{equation}\label{dualgrav}
ds^2= f(u)\big[-h(u)dt^2 + dx^2 + dy^2 +  dz^2 + {du^2\over h(u)}\big],
\end{equation}
\begin{equation}
{\rm where}, f(u)= {l^2\over u^2} \quad {\rm and}\quad h(u)= \frac{u^2}{l^2} + \frac{1}{4 \bar \alpha}\Big(1-\sqrt{1-8\bar\alpha + \frac{32 \bar \alpha M u^4}{l^6} + \frac{16 \bar a \bar \alpha u^3}{3 l^3}}\Big)\nonumber
\end{equation}
and the radius of the horizon, $u_h$, can be obtained by solving the equation,
\begin{eqnarray}
 h(u_{h})= \frac{u_h^2}{l^2} + \frac{1}{4 \bar \alpha}\Big(1-\sqrt{1-8\bar\alpha + \frac{32 \bar \alpha M u_h^4}{l^6} + \frac{16 \bar a \bar \alpha u_h^3}{3 l^3}}\Big) = 0.
\label{hor}
\end{eqnarray}
We will consider, the end points of the open string corresponding to $Q\bar Q$ pair are located at $x = \pm \frac{L}{2}$ respectively and is elongated in the bulk space time with the symmetry around $x = 0$.

With this background, we evaluate the Nambu-Goto action, (\ref{ng}). We choose: $\sigma^0 = t,\,\, \sigma^1 = x$ (static gauge) and the metric induced on the worldsheet, given by $h_{\beta \gamma} = \partial_\beta X^\mu\partial_\gamma X^\nu g_{\mu\nu}$ becomes,
\begin{equation}
ds^2 = f(u)\big[-h(u)dt^2 + \big\{1 + \frac{u'^2}{h(u)}\big\}dx^2\big].
\end{equation}
After $t$ integration is carried out, the Nambu-Goto action becomes
\begin{equation}\label{ng2}
S_{NG} = - \frac{{\mathcal T}}{2 \pi \alpha^\prime } \int\limits_{-L/2}^{L/2} dx f(u) \sqrt{h(u) \left( 1 +\frac{ {u^\prime}^2}{h(u)} \right)}.
\end{equation}
The equations of motion for the embedding coordinate $u(x)$, ensuing from the above action is given by
\begin{equation}\label{ng3}
u^\prime (x) = - \sqrt{ h(u)  \left( \frac{f(u)^2 h(u)}{f(u_0)^2 h(u_0)} - 1 \right)},
\end{equation}
implying the string is extended towards the horizon up to a turning point $u=u_0$ and goes back to the boundary in a symmetric manner with $u(0)=u_0$.
The distance between the two end points of the string can be obtained as
\begin{equation}\label{qqseparation}
L = \int\limits_{- L/2}^{L/2} dx = 2 \int\limits_0^{u_0} \frac{du}{u'} =2 \int\limits_0^{u_0}du \left[ h(u)
\left(  \frac{f(u)^2 h(u)}{f(u_0)^2 h(u_0)} -1 \right) \right]^{-\frac{1}{2}},
\end{equation}
where $u_0$ is the upper bound on the location of the tip of the string that is extended towards horizon.

\begin{figure}[h]
\begin{center}
	\mbox{
	\subfigure[]{\includegraphics[width=7.5cm]{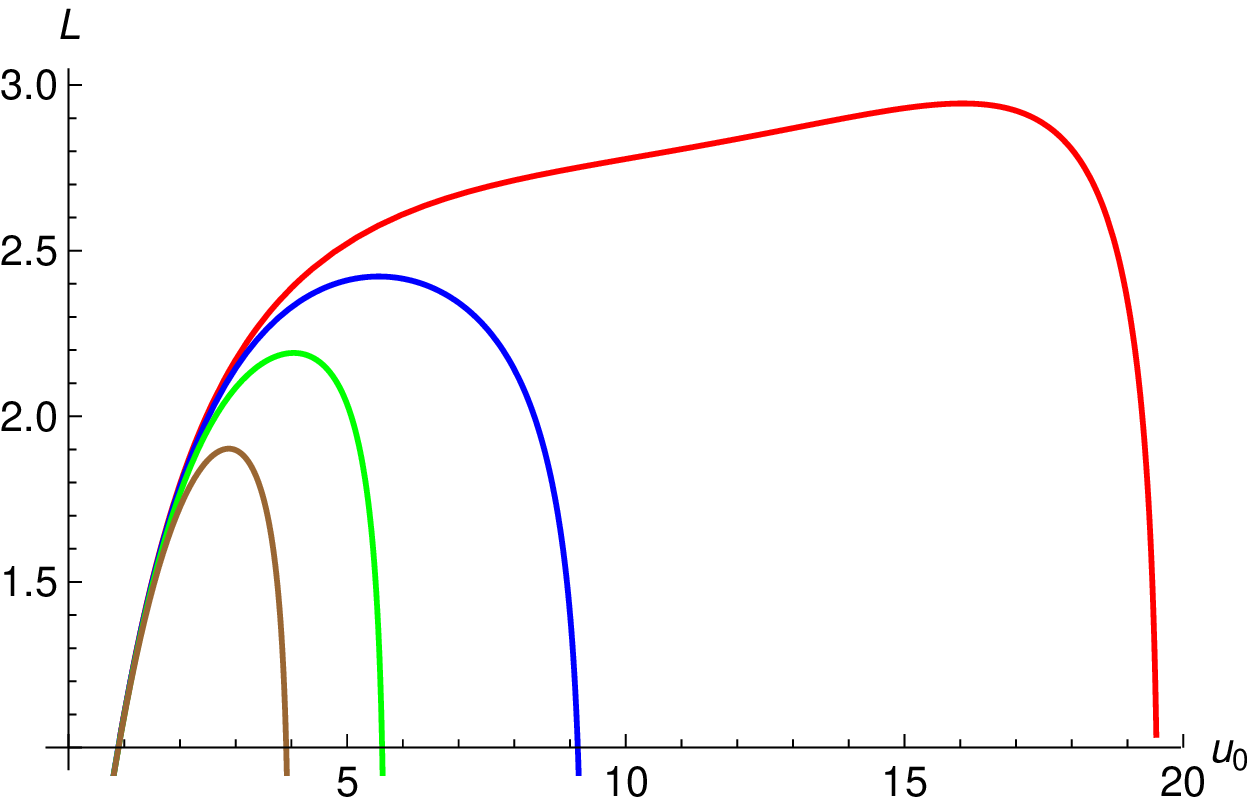}}
\quad
	\subfigure[]{\includegraphics[width=7.5cm]{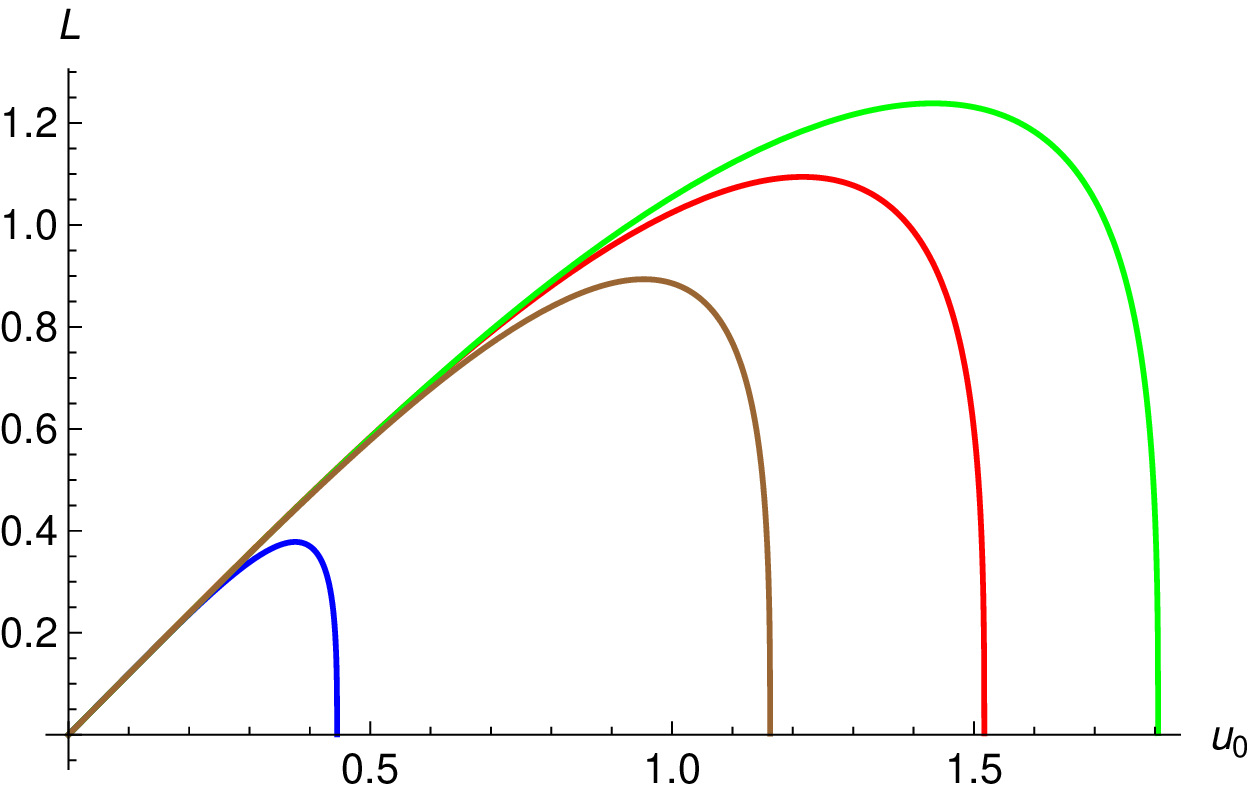}}}
\end{center}
\caption{Q$\bar Q$ -distance vs. $u_0$: $\bar a=.1, \bar{ \alpha}=.001$, $T=T_1$ (Red), $T=T_2$ (Blue); $\bar a=.2,\alpha=.004$, $T=T_1$ (Green), $T=T_2$ (Brown); black holes with radius (a) small (b)  large.}
\label{distance}
\end{figure}

 We have plotted the quark-antiquark separation $L$  with respect to $u_0$, the upper bound on the location of the tip of the string extended towards the horizon in figure \ref{distance}. One can observe from the figure, as the tip located at $u_0$ gets closer to horizon,  quark-antiquak separation increases. It becomes maximum when $u_0$ gets extremely near to the horizon and once it touches the horizon, $L$ becomes zero. This fact can be interpreted as a  breakdown of the U-shaped string configuration to two parallel straight string configuration which corresponds to unbound state of $Q\bar Q $ pair. This happens in case of the small as well as the large black hole. It suggests that confined state might not be present in the dual theory living on the boundary, as expected in gravity analysis.

\section{Quark-antiquark potential}\label{potential}
In the present section we are interested to study free energy associated with the heavy quark-antiquark pair of the dual theory living on the boundary. The Wegner-Wilson loop encodes the free energy associated with a quark-antiquark pair \cite{Ewerz:2016zsx},
\begin{equation}\label{wwloop}
\langle W({\mathcal C}_{L,{\mathcal T}})\rangle \sim \exp[ -i F_{Q{\bar Q}}(L){\mathcal  T} ],\quad {\mathcal T}\rightarrow\infty,
\end{equation}
where $F_{Q{\bar Q}}(L)$ is the $Q\bar{Q}$ free energy. It depends on the temperature as the expectation value is  for a thermal state. Since in the present discussion we will consider the range of $L$ only up to a screening distance $L_s$, the free energy $F_{Q\bar{Q}}$ will agree with the quark-antiquark potential $V_{Q\bar{Q}}$ in our case.

According to holographic principle, Nambu-Goto action of an open string given in equation (\ref{ng}) is related to the Wegner-Wilson loop where ${\mathcal C}$, the contour of integration is identified with open string worldsheet boundary. The on-shell string action then is related to expectation value,
\begin{equation}\label{wwloop1}
\langle W({\mathcal C})\rangle \sim \exp[ i S_{NG}({\mathcal  C} )],
\end{equation}
where $S_{NG}({\mathcal  C}) $ is the on-shell Nambu-Goto action of the string whose expression is given in (\ref{ng2}). Comparing (\ref{wwloop}) and
(\ref{wwloop1}), we obtain
\begin{equation}
F_{Q{\bar Q}}(L) \sim - \frac{S_{NG}({\mathcal  C})}{{\mathcal T}}.
\end{equation}
$F_{Q{\bar Q}}(L)$, free energy for quark-antiquark can be obtained by
substituting (\ref{ng3}) in (\ref{ng2}).
As has been discussed elaborately in \cite{Ewerz:2016zsx}, computation of  $S_{NG}$ in general gives rise to divergences and this expression needs to be renormalised.  We  follow the same renormalisation prescription as given in \cite{Ewerz:2016zsx} and obtain a renormalised expression for the quark-antiquark free energy, given by
\begin{equation}\label{FQQ1}
\frac{\pi F_{Q\bar{Q}}(u_0)}{\sqrt{\lambda}} = \int\limits_0^{u_0} du \left[ f(u)
\sqrt{ \frac{f(u)^2 h(u)}{  f(u)^2 h(u) - f(u_0)^2 h(u_0)} } - \frac{1}{u^2} \right] - \frac{1}{u_0} .
\end{equation}
Usually quark-antiquark binding energy can be obtained from U-shaped string action with action of two straight strings from boundary to horizon subtracted from it, which is given by \cite{Ewerz:2016zsx}
\begin{equation}\label{EQQ}
\frac{\pi E_{Q\bar{Q}}(u_0)}{\sqrt{\lambda}} = \int\limits_0^{u_0} du
\left[
f(u) \sqrt{
				 \frac{f(u)^2 h(u)}{ f(u)^2 h(u) - f(u_0)^2 h(u_0)}
				  } - 1
\right] - \int\limits_{u_0}^{u_h} f(u) du .
\end{equation}

In what follows, in both the cases of small and large black hole we can obtain free energy and binding energy for the quark-antiquark potential. We begin with black holes that has been obtained without the higher derivative coupling. Some result has already appeared in \cite{tanay1} and here we have a more elaborate analysis. If we drop the higher derivative terms, metric functions are \cite{tanay1}
\begin{equation}
f(u) = \frac{1}{u^2},\quad\quad
h(u) = 1 - \Big(\frac{u}{u_h}\Big)^4 + u^2 \Big[ 1 - \Big(\frac{u}{u_h}\Big)^2\Big] - \frac{2\bar{a}}{3} u^3 \Big(1 - \frac{u}{u_h}\Big),
\end{equation}
where we have traded $M$ for horizon radius $u_h$. Here and in the rest of the calculation, we set $l=1$ 

We have substituted these expressions in the general formula for the free energy and the binding energy for the quark-antiquark pair derived in (\ref{FQQ1}) and (\ref{EQQ}) respectively. Since the integration cannot be carried out to obtain expressions in terms of known analytic functions, we have evaluated them numerically. We obtain the quark-antiquark separation from the expression (\ref{qqseparation}) as a function of $u_0$,  the closest approach of the string worldsheet towards the horizon. It turns out that as $u_0$ varies there is a maximum value of the quark-antiquark separation, which depends on the temperature as well.

\begin{figure}[h]
\begin{center}
\mbox{
\subfigure[]{\includegraphics[width=7.5 cm]{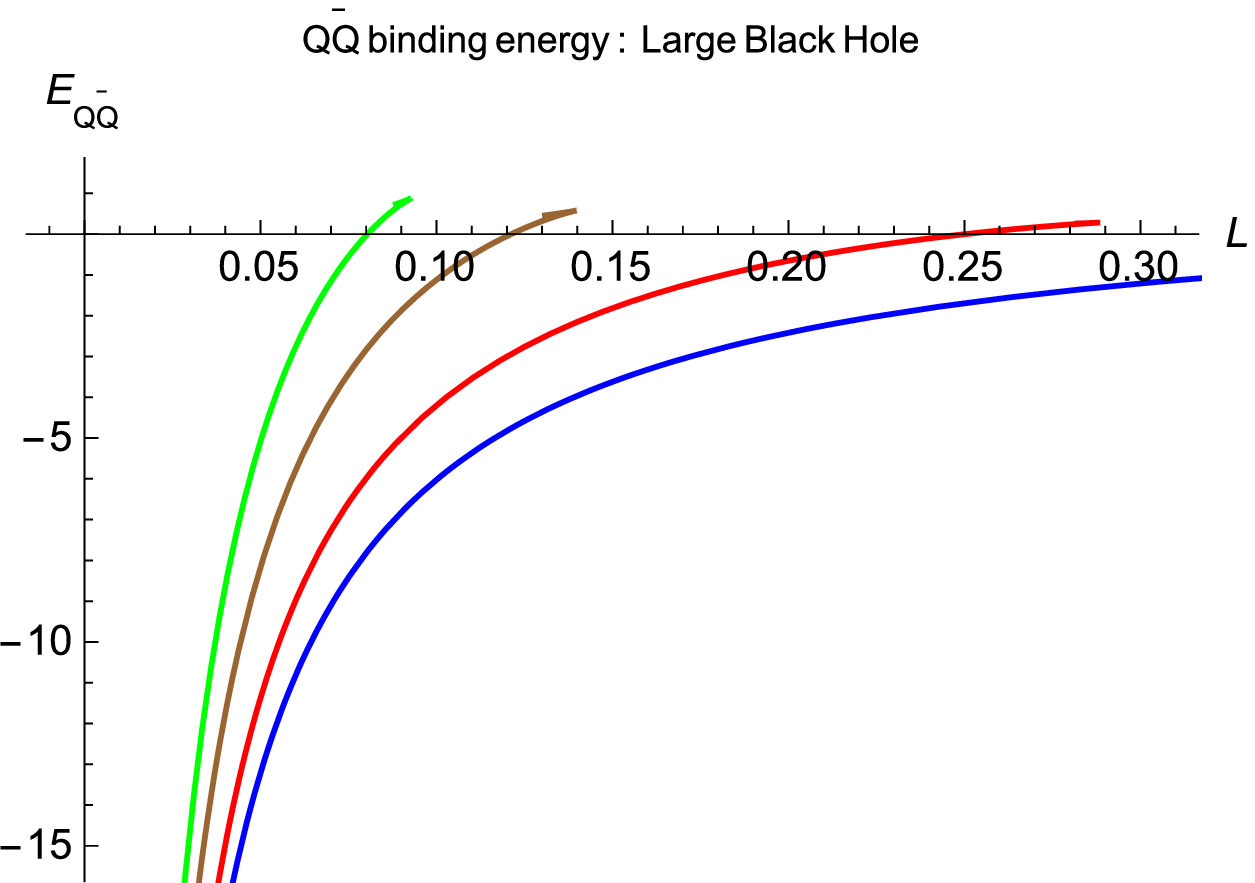}}
\quad
\subfigure[]{\includegraphics[width=7.5 cm]{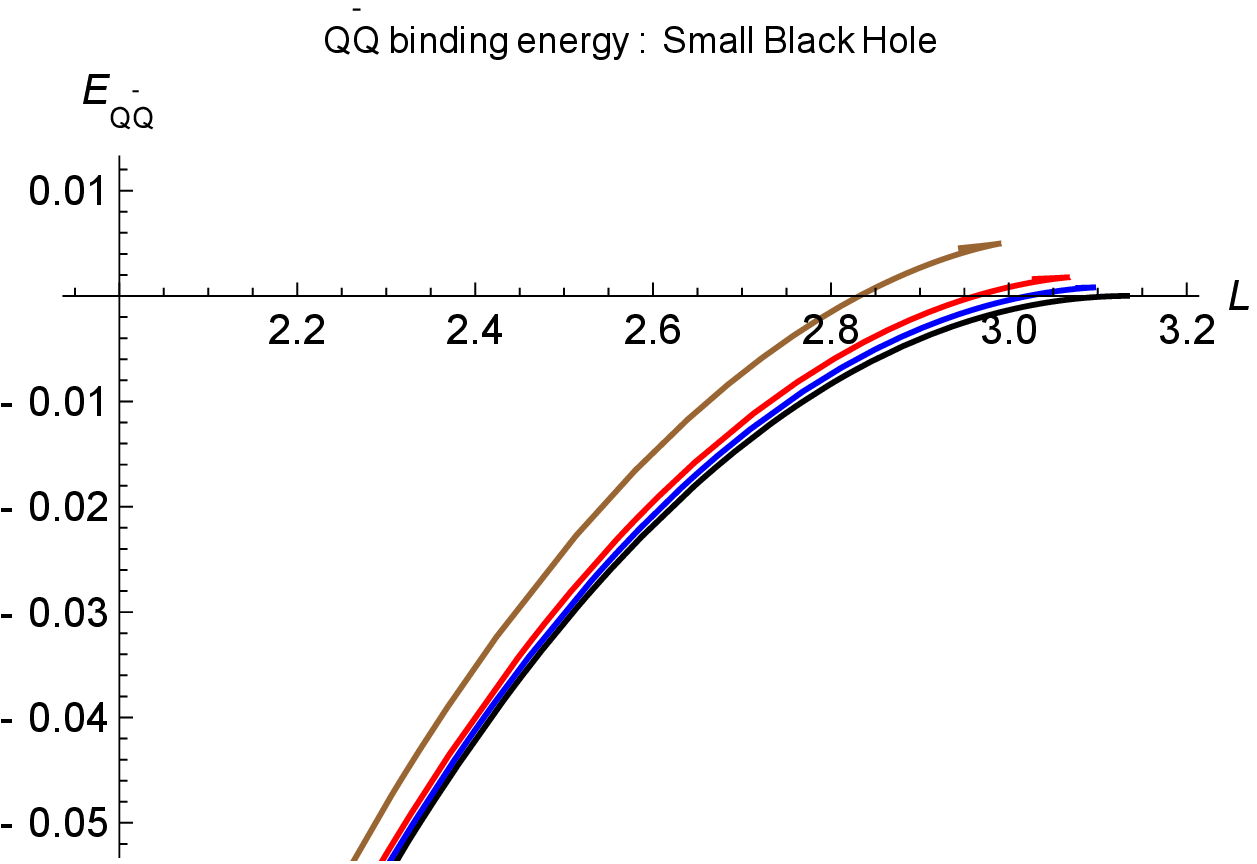}}}
\caption{${\bar a} =.05$; (a) $Q\bar{Q}$ binding energy for the large black hole for temperatures, T= 3 (Green), 2 (Brown), 1 (Red), 0.5 (Blue)
(b) $Q\bar{Q}$ binding energy for the small black hole for temperatures, T=  2 (Brown), 1 (Red), 0.5 (Blue), .01 (Black);}
\label{EQQf}
\end{center}
\end{figure}

\begin{figure}[h]
\begin{center}
\mbox{\subfigure[]{\includegraphics[width=7.5 cm]{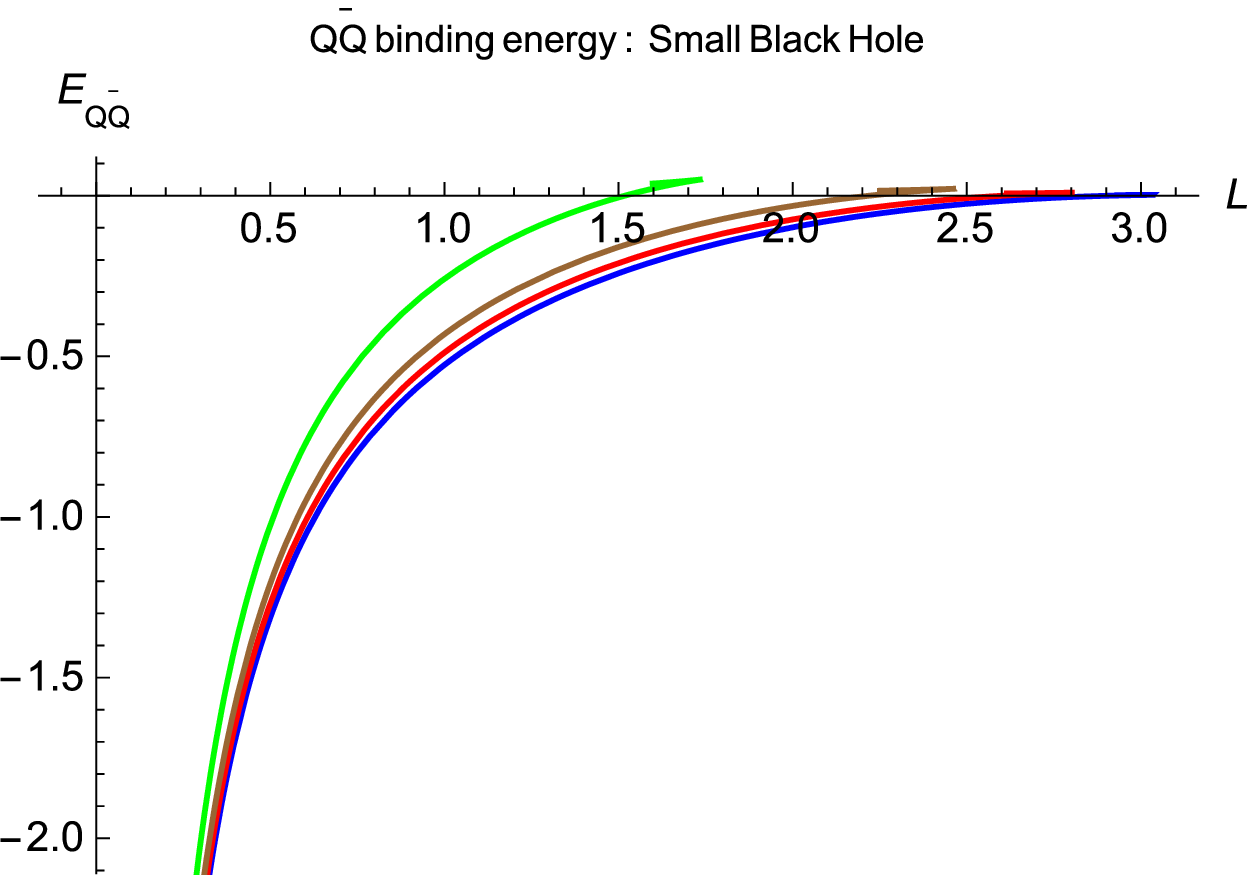}}
\quad
\subfigure[]{\includegraphics[width=7.5 cm]{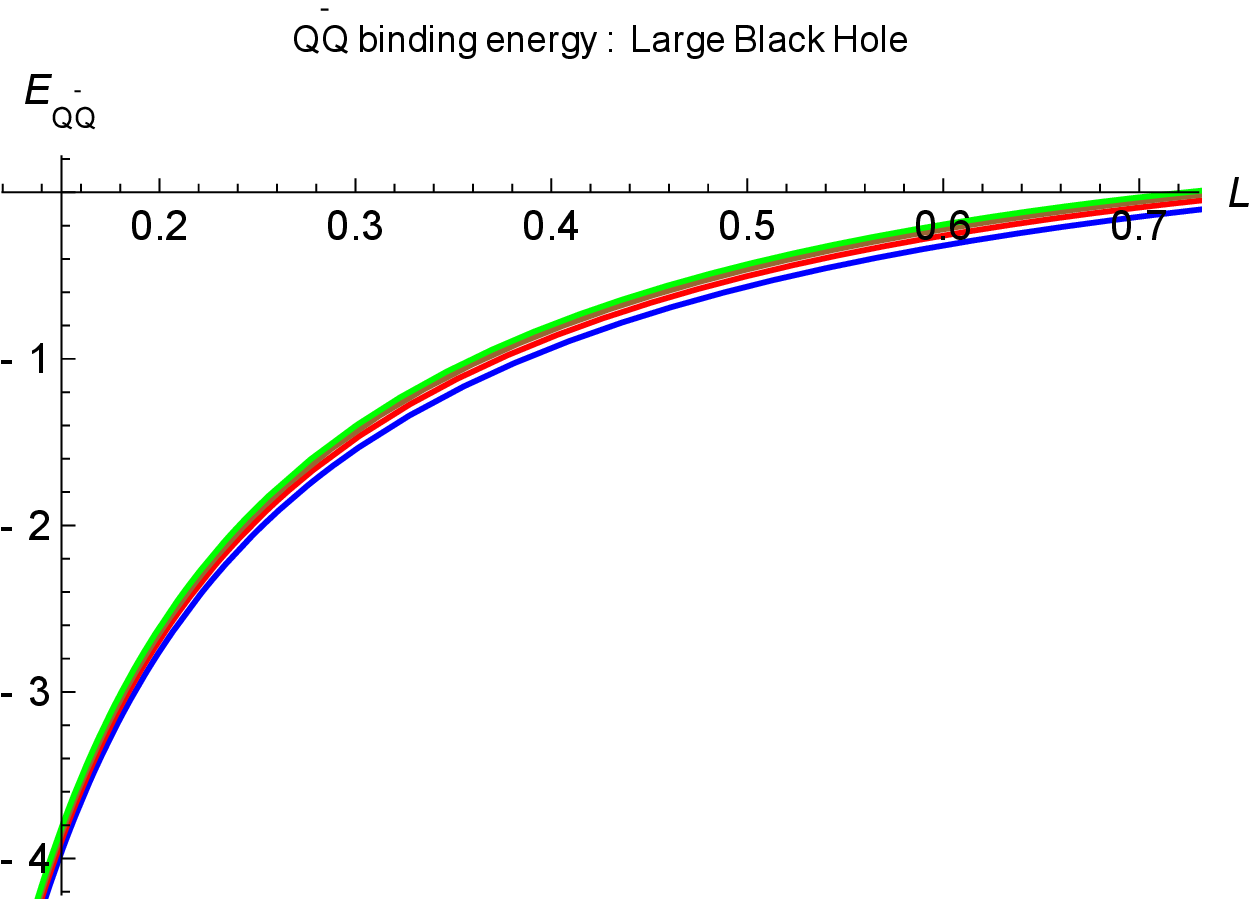}}}
\caption{$Q\bar{Q}$ binding energy for different ${\bar a}$ values, ${\bar a}$= .4 (Green),
.3 (Brown), .2 (Red), .1 (Blue) (a) Small black hole,  ${T} =1$ (b) Large black hole, ${T} = 0.4$;}
\label{FQQ}
\end{center}
\end{figure}

In order to find out the thermodynamically stable quark-antiquark bound states we have plotted the binding energy  versus the quark-antiquark separation for both the black holes in the figure \ref{EQQf}. One may observe that the binding energy of the bound state for large black hole is smaller than the small black hole background indicating that the bound states are more stable in the background of large black hole. One can also notice that for both the cases, after a certain value of separation distance the binding energy of the pair is getting positive value. Therefore the bound state exist up to an upper value of separation distance which is called screening length $ L_s$ and that is finite. The screening length decreases with the increase of temperature and bound state becomes more unstable at higher temperature. We compare the screening length for small and large black hole background and find screening length is larger in small black hole configuration showing it admits larger separation, but in none of the cases, it can go to infinity or can be interpreted as deconfined phase. In figure \ref{FQQ} we have analysed the same features for different values of string cloud density by keeping temperature fixed. The qualitative behaviour remains unchanged and here string cloud density plays a role analogous to the temperature. Finally we can conclude that deconfined state of quark and antiquark is the stable configuration.

\begin{figure}[h]
\begin{center}
\mbox{\subfigure[]{\includegraphics[width= 7.5cm]{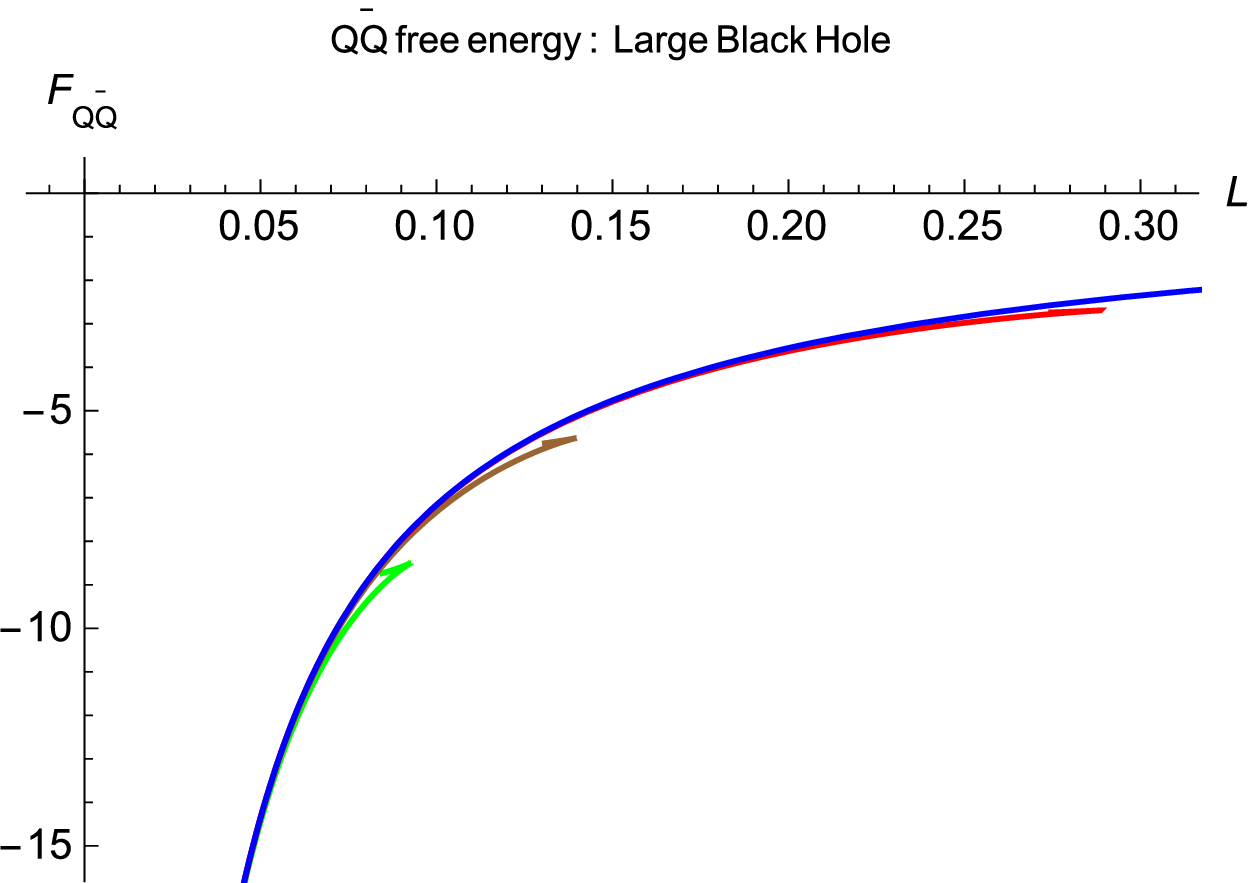}}
\quad
\subfigure[]{\includegraphics[width=7.5 cm]{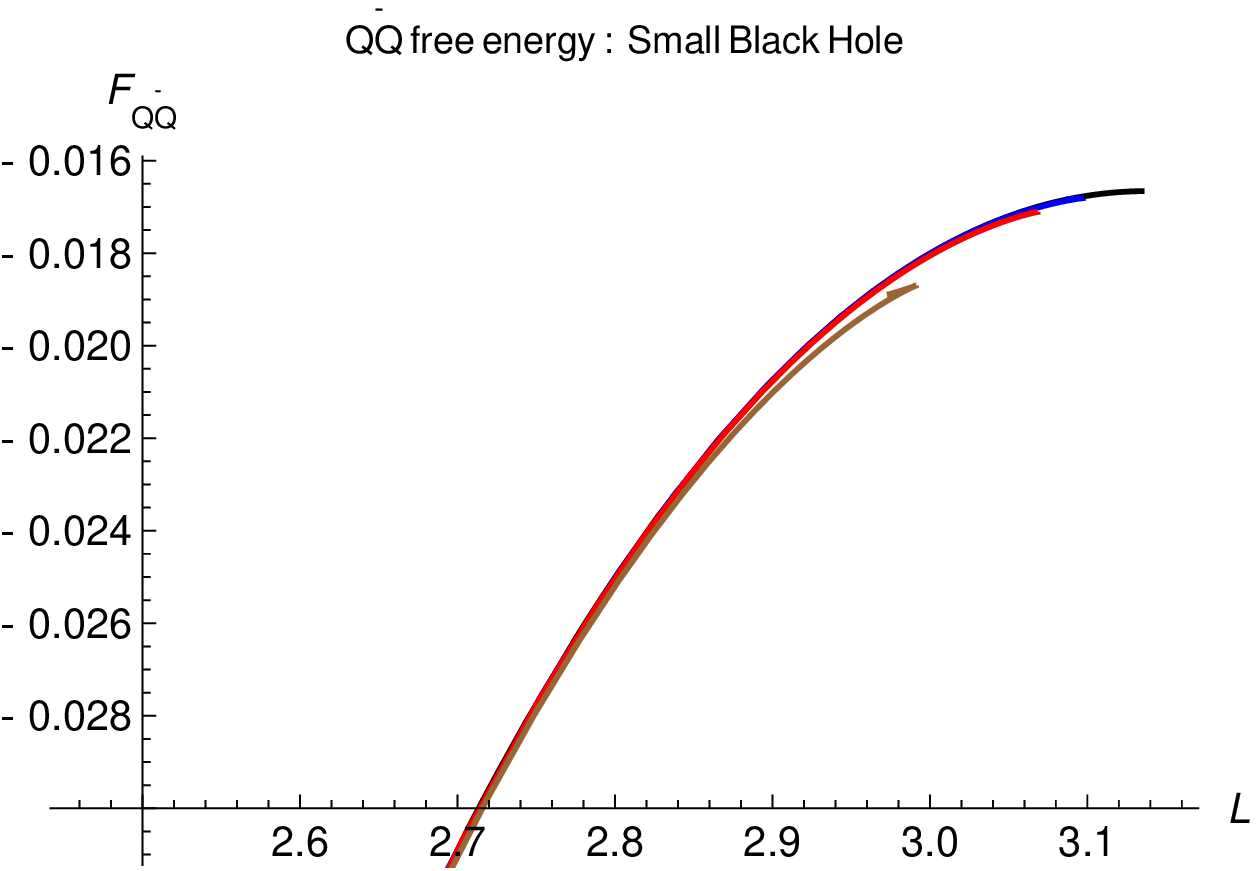}}}
\caption{${\bar a} =.05$; $Q\bar{Q}$ free energy vs. L: (a) large black hole with T= 3 (Green), 2 (Brown), 1 (Red), 0.5 (Blue)
(b) small black hole with T=  2 (Brown), 1 (Red), 0.5 (Blue), .01 (Black);}
\label{FQQf}
\end{center}
\end{figure}

The quark-antiquark free energy obtained from (\ref{FQQ1}) are numerically evaluated for both the black holes in the same range of temperatures and are plotted with respect to the quark-antiquark separation in figure \ref{FQQf}. Comparing with the binding energy plotted in figure \ref{EQQf}, one can observe that with variation of the temperature the free energy associated with quark-antiquark pair does not vary much, though there is a reasonable difference in the binding energy, mostly arising from the subtraction of the energy due to the straight strings.

Modification due to higher derivative terms can be studied by considering black hole solution obtained from Einstein-Gauss-Bonnet action as discussed in the previous sections. Once higher derivative terms are taken into account, the metric is given in (\ref{genmet}) and (\ref{comp}).  In terms of the general form of metric given in (\ref{dualgrav}) the metric functions become
\begin{equation}\begin{split}
f(u) &= \frac{1}{u^2},\\
h(u) &= u^2 + \frac{1}{4 \bar\alpha} \left[ 1 - \sqrt{ 1 - 8\bar\alpha \Big(1 - \frac{u^4}{u_h^4}\Big) + \frac{16 \bar a \bar\alpha }{3} u^3 \Big( 1 - \frac{u}{u_h}\Big) + 32 \bar\alpha u^4 \Big(\frac{1}{4 u_h^2} +\frac{\bar\alpha}{2}\Big)}\quad \right],\end{split}
\end{equation}
where, as usual, we have traded mass parameter $M$ for the horizon radius $u_h$.

\begin{figure}[h]
\begin{center}
\mbox{
\subfigure[]{\includegraphics[width=7.5 cm]{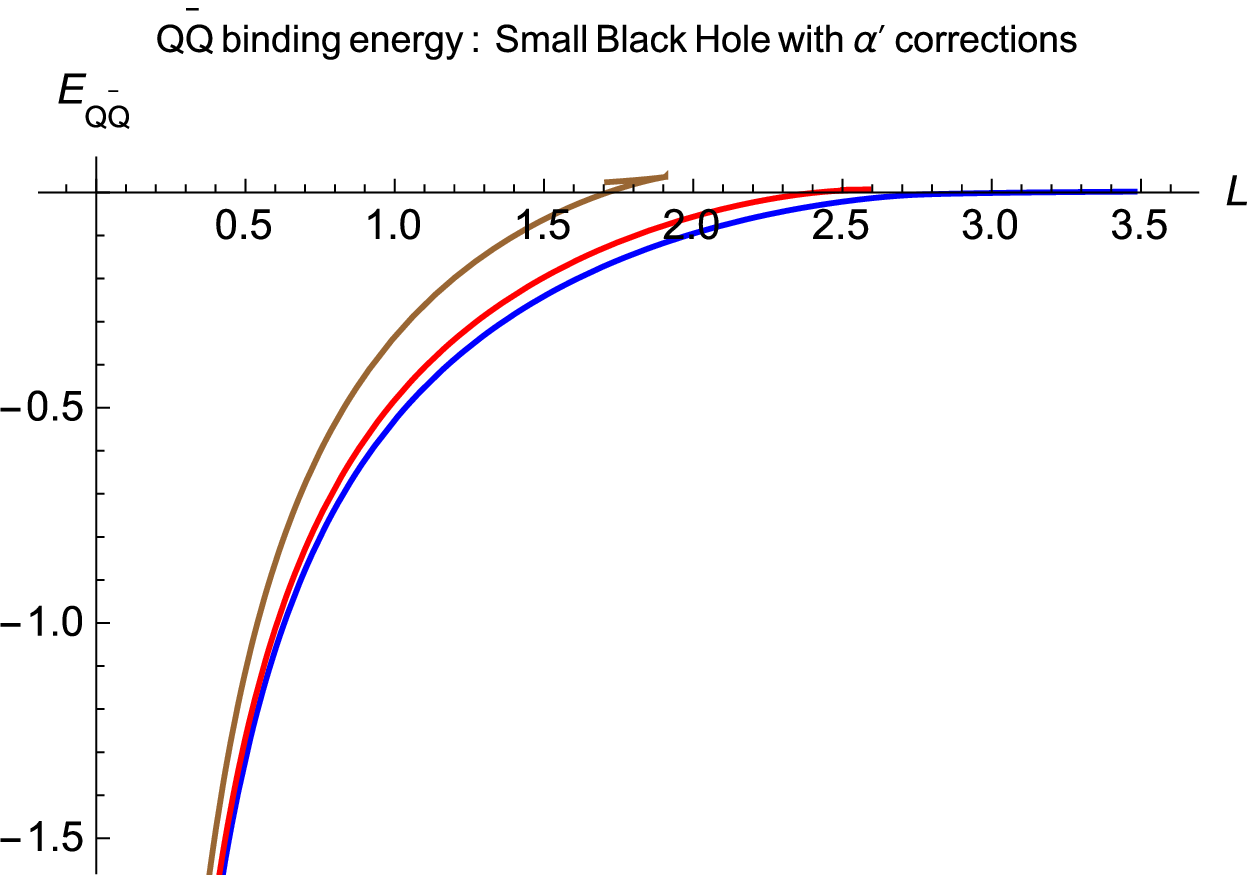}}
\quad
\subfigure[]{\includegraphics[width=7.5 cm]{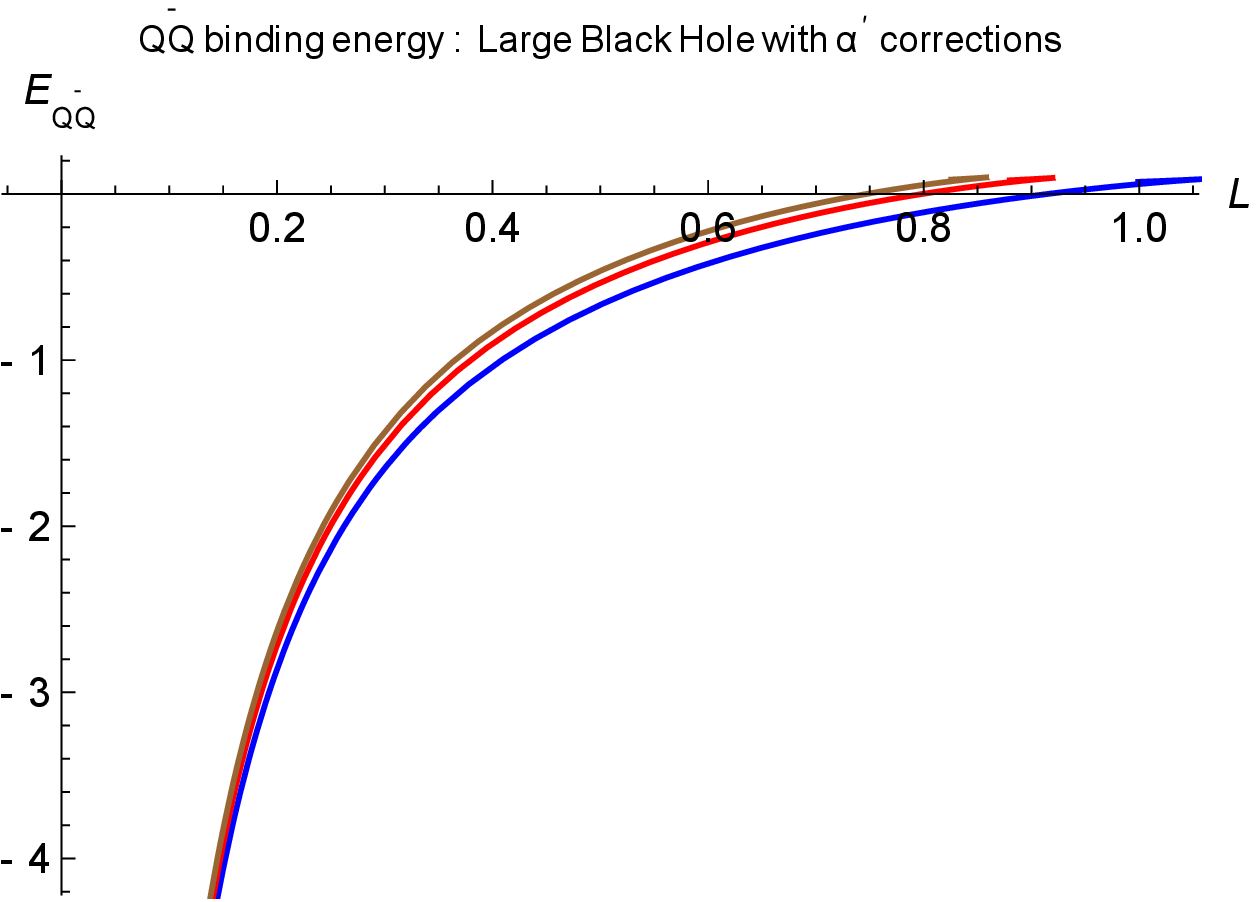}}}
\caption{ $Q\bar{Q}$ binding energy vs. L for ${\bar a} =.05$; $T=.441$; $\bar{\alpha}$ = .001(Blue), .004(Red), .0082(Brown)
(a) small black hole (b) large black hole}
\label{FQQh}
\end{center}
\end{figure}

Substituting the above expressions in the general formula for the binding energy and the separation for the quark-antiquark pair given in (\ref{FQQ1}) and (\ref{qqseparation}) respectively, we have plotted the binding energy against $L$. We consider four different values of $\bar\alpha$, the coefficient of higher derivative coupling in the plots for both the black holes as given in figure \ref{FQQh}.

Since, as the $\bar\alpha$ varies, the critical temperature  keeps on changing, we have chosen suitable values of the temperatures for the two different black holes. As the plots show, with the increase of the coefficient of the higher derivative terms the binding energy increases making the quark-antiquark bound state less stable. Furthermore, for a fixed amount of binding energy the length of the maximum separation between the quark and the antiquark decreases. This indicates, with addition of subleading corrections in `t Hooft coupling, the $Q\bar{Q}$ bound states will become loosely bound favouring deconfinement. Both the small and the large black holes share this feature. Though we have given the plot only for a single temperature in each case, this feature qualitatively persists with variation of the temperature.

\begin{figure}[h]
\begin{center}
\mbox{
\subfigure[]{\includegraphics[width=7.5cm]{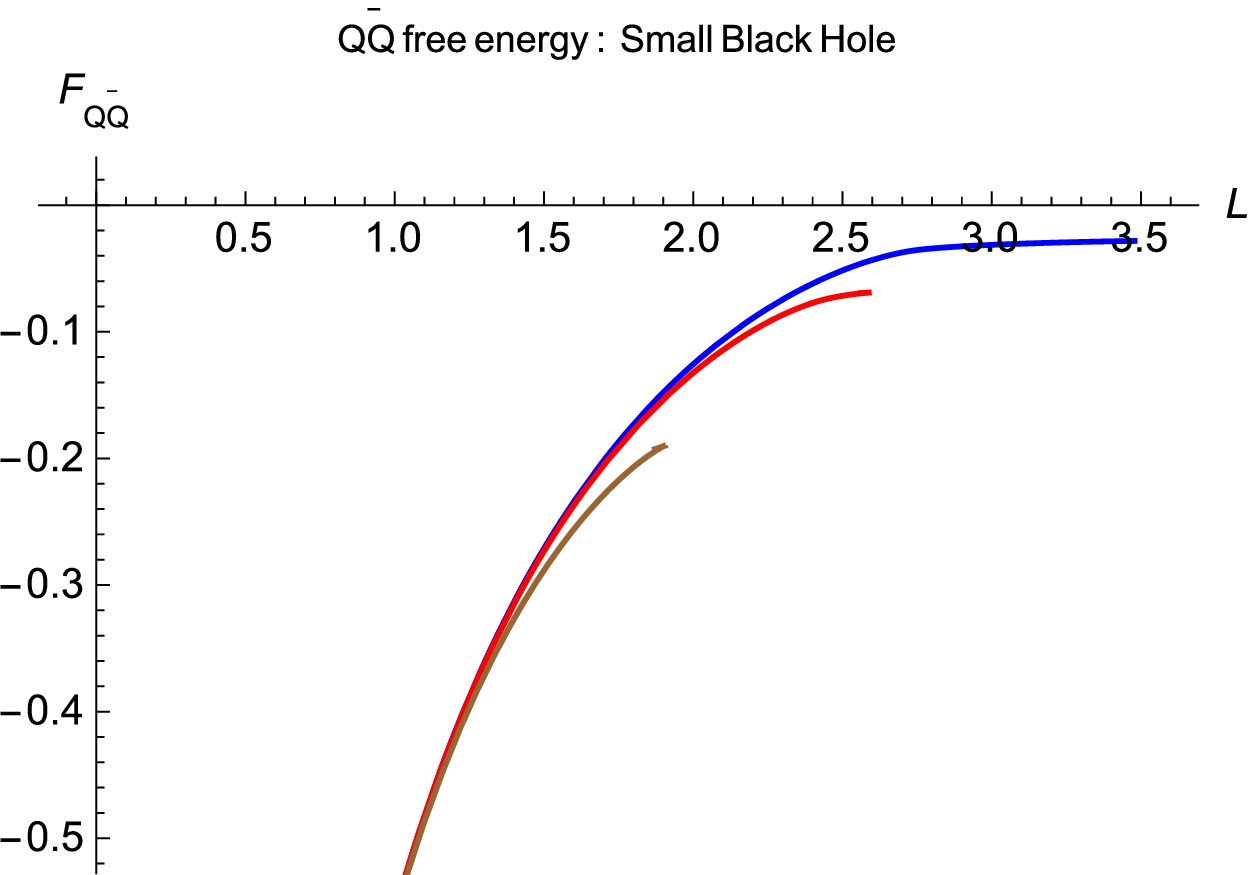}}
\quad
\subfigure[]{\includegraphics[width=7.5 cm]{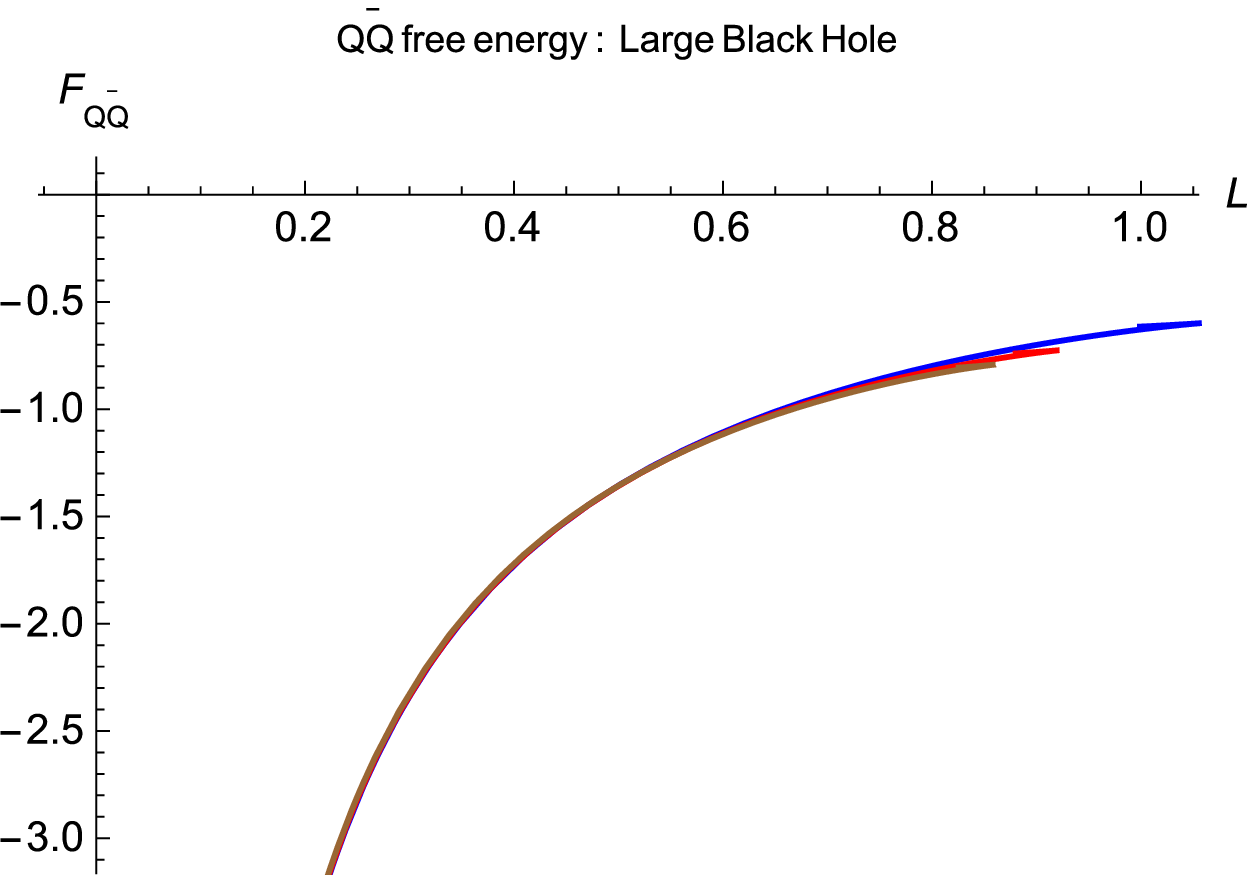}}}
\caption{$Q\bar{Q}$ free energy vs. L for ${\bar a} =.05$; $T=.441$; $\bar{\alpha}$ = .001(Blue), .004(Red), .0082(Brown)
(a) small black hole (b) large black hole}
\label{FQQhf}
\end{center}
\end{figure}

At last we have studied the quark-antiquark free energy for both the black holes. We plot the analysis in figure \ref{FQQhf} and observe that qualitative features are same as without higher derivative term.

\section{Summary}\label{conclude}

We have considered solutions in the Einstein Gauss-Bonnet gravity along with an
external string clouds. It is found that the theory admits three different black holes, dubbed as small, medium and large depending on the horizon radii of the respective black holes. As the density of string cloud increases and/or the higher derivative correction dominates, the parameter region for which one gets the three solutions shrinks. Beyond the critical values of string cloud density or higher derivative correction, the theory admits only a single solution.

The analysis of the specific heat associated with the solutions implies that the medium size black hole is unstable, while the other two solutions are thermodynamically stable. From the free energy consideration we find at a higher temperature the thermodynamically faouvored configuration is the black hole with largest horizon radius. We have analysed Landau function to examine possible transitions among these solutions.

Using holographic duality, we examine the quark-antiquark bound states. We find such bound states exist up to a distance of screening length, beyond which they get separated, indicating the dual theory is in a deconfined phase. The screening length turns out to be larger for small black hole than that in large black hole, though it is finite and cannot be extended indefinitely as happened in a confined phase. From the study of quark antiquark binding energy, we find it decreases as the string cloud density and the higher derivative correction becomes larger. This may be interpreted as, in presence of large number of quarks in the background, the quark-antiquark bound states will be loosely bound. A similar effect appears as the subleading corrections in `t Hooft coupling becomes increasingly dominant.

It may be interesting to consider further extension of this model. In particular one can consider the charged black hole solution in presence of string clouds and examine the different phases admitted by such theory.


\end{document}